\documentclass[aps,amsmath,amssymb,showpacs,showkeys]{revtex4-2}
\usepackage[dvips]{graphicx}
\usepackage{times}
\usepackage{multirow}
\usepackage{xcolor}
\usepackage{orcidlink}
\usepackage{hyperref}
\hypersetup{
  colorlinks=true,
  urlcolor=blue,
  linkcolor=red,
  citecolor=blue
}

\begin{document}
\title{Quasinormal modes, temperatures and greybody factors of black holes in 
a generalized Rastall gravity theory}

\author{Ronit Karmakar  \orcidlink{0000-0002-9531-7435}}
\email[Email: ]{ronit.karmakar622@gmail.com}

\affiliation{Department of Physics, Dibrugarh University,
Dibrugarh 786004, Assam, India}

\author{Umananda Dev Goswami  \orcidlink{0000-0003-0012-7549}}
\email[Email: ]{umananda2@gmail.com}

\affiliation{Department of Physics, Dibrugarh University,
Dibrugarh 786004, Assam, India}

\begin{abstract}
We introduce a modification in the energy-momentum conservation violating 
Rastall's theory of gravity and obtain a Reissner-Nordstr\"om-type black hole 
solution in spacetime surrounded by a cloud of strings and charge fields. 
We examine the horizons of the black hole along with the influence of the 
parameters of the model on it. The scalar quasinormal modes (QNMs) of 
oscillations of the black hole are also computed using the 6th order WKB 
approximation method. It is seen that the Rastall parameter $\beta$ and the
newly introduced energy-momentum tensor trace parameter $\alpha$ as well as 
the charge parameter $q$ and strings field parameter $a$ influence the 
amplitude and damping of the QNMs. From the metric function, we obtain the 
temperature of the black hole and study the effects of the four model 
parameters $\beta$, $\alpha$, $q$ and $a$ on the temperature. We then examine 
the greybody factors associated with the black hole and the corresponding total 
absorption cross-section for it. It is seen that the modification 
we introduced in the Rastall theory has a drastic effect on various properties 
of the black hole and may lead to interesting outcomes in future when 
better detection techniques will be available with the LISA and the Einstein 
Telescope. 

\end{abstract}


\maketitle
\section{Introduction}
The theory of General Relativity (GR) was endowed with the revolutionary 
description of gravity, which undoubtedly has appeared as the major 
milestone in the field of modern astrophysics and cosmology. The two of the 
most significant predictions made by GR have been observationally verified 
most recently: the detection of gravitational waves (GWs) from the binary black 
hole system merger by the LIGO-Virgo collaboration \cite{1,2,3,4,5,6} and the 
first images of the black hole M87* by the Event Horizon Telescope (EHT) 
\cite{7,8,9,10,11,12}. These two observational verifications increase the 
importance of the theory even today. Moreover, GR has been tested in the 
post-Newtonian levels to a high precession viz., by the light deflection, 
Shapiro time delay and perihelion advance of Mercury \cite{12-1,13,42-3}. 
Confirmation of the validity of GR has also been inferred from the Hulse-Taylor 
binary pulsar timing array data which matched the GR predicted GWs damping to 
high accuracy \cite{13-1,14,15}. All these reasons are more than sufficient 
to state that GR is indeed a successful theory of gravity. 

However, GR is afflicted with some issues. The theory is non-renormalizable in 
the high energy regime \cite{16}. Also in the infrared regime, there are 
deviations from experimental findings like the accelerated expansion of 
the Universe \cite{17,18} and the dark matter sector \cite{19}. That is, we 
cannot explain the accelerated expansion of the Universe and features that 
indicate the hidden matter content of the Universe with this theory. To 
overcome these issues, many theories of gravity have been proposed, the most 
common class of which are the Modified Theories of Gravity (MTGs) (see 
\cite{20,21} and references therein). In these theories, either the matter 
side or the curvature side of the general Einstein field equations are 
modified to mitigate these issues. $\Lambda$CDM model \cite{22} represents 
the simplest amongst the matter-modified theories or the usually known dark 
energy models \cite{22-01}, and  $f(R)$ theory \cite{22-1} represents one 
of the simplest among the spacetime geometry-modified theories or the commonly
referred MTGs. Some other MTGs include Rastall gravity 
\cite{23,24,25,26,27,28,28-1}, $f(R,T)$ gravity \cite{29}, $f(Q)$ gravity 
\cite{30,31,32,32-1,32-2,32-3,32-4}, $f(R,L_M)$ gravity \cite{33} etc.\\ 
\indent
In 1972, P.\ Rastall proposed his theory of gravity which gained popularity 
eventually, as this theory was capable of predicting observational results 
like galactic rotation curves \cite{34} and accelerated expansion of the 
Universe \cite{35}. However, the point to be noted is that this theory is not 
derived from an action principle approach which is one of the drawbacks of the 
theory. In this theory, the energy-momentum conservation law is modified and 
the covariant derivative of the energy-momentum tensor is taken to be 
proportional to the covariant derivative of the Ricci scalar $R$. This theory 
recovers GR in the regime of zero background curvature. Thus, Rastall's theory 
is a generalization of GR in a sense. However, Visser in his paper 
\cite{28} argued that Rastall gravity (RG) is equivalent to GR and that RG 
does not provide any new insights compared to GR. A number of counterarguments 
have been advanced which restate that RG is a more general theory of gravity 
and it includes GR as a special case. In Ref.\ \cite{28-new1}, the authors 
stated clearly that any metric theory including $f(R)$ gravity can be 
transformed into a GR-like form but that does not imply that the theories are 
equivalent. In fact, they provide compelling arguments in support of RG. 
Moreover, in Ref.\ \cite{28-new2}, Darabi et al.\ have pointed out deviations 
of RG from GR based on cosmological observations. Similarly, in 
Ref.\ \cite{28-new3}, authors investigate compact stellar objects using 
modified Rastall teleparallel gravity. Recently, Hansraj et 
al.\ \cite{28-new4} reinforced the claim of Darabi and his team. 
Moreover this theory provides predictions regarding the age of the Universe 
\cite{36} and the Hubble parameter \cite{36}. This theory can justify the 
gravitational lensing process \cite{37} and also the existence of traversable 
wormhole solutions has been shown recently \cite{38}. In one work, the 
authors, in the context of Rastall gravity, studied the gravitational collapse 
of a homogeneous perfect fluid \cite{39}. In another recent work, the authors 
studied ABG-type black holes in Rastall gravity surrounded by a string's cloud 
in the presence of non-linear electrodynamic sources \cite{26}. In the 
paper \cite{39-1} the authors studied the neutron star for a 
realistic equation of state under the framework of Rastall gravity \cite{39-1}. 
Similarly, the authors of the paper \cite{39-2} have studied the thermodynamic 
properties along with Joule-Thomson expansion and the optical properties of 
the black hole surrounded by quintessence, under the framework of Rastall 
gravity. In Ref.\ \cite{40} authors introduced a generalised form of the 
Rastall gravity and studied the compact objects in this framework. 
The motivation to choose Rastall theory over other modified gravity 
theories comes from the fact of the simplicity of field equations of RG as 
compared to other theories. Moreover, RG is capable of handling modern 
observational constraints and various avenues of theoretical research have 
been persued in recent times considering this framework.\\[2pt]
\indent
Black holes are frequently studied with a surrounding field that impacts their 
properties to a good extent. In 2003, Kiselev studied black holes surrounded 
by the quintessence \cite{41} and since then many research works have been 
published with such fields surrounding the black holes 
\cite{25,26,27,42,42-1,42-2,42-4,42-5,43,44,44-0,44-1}. P.\ 
Letelier studied black holes surrounded by 
clouds of strings \cite{44}. Recently, different surrounding fields like dark 
matter fields or dark energy fields have been considered. Heydarzade and 
Darabi considered the Kiselev-like charged/uncharged black hole solutions 
surrounded by a perfect fluid in the framework of Rastall gravity \cite{25}. 
In another work, the authors implemented GUP corrections into the black hole solution 
surrounded by quintessence matter \cite{42}. They studied the QNMs and 
thermodynamic properties of the black hole. Chen et.\ al.\ 
\cite{44-1} studied the Hawking radiation of a d-dimensional black hole 
surrounded by quintessence fields. The existence of Nariai black holes for 
some specific parameters has been shown in Ref.\ \cite{44-2}. Thermodynamics of 
the quantum-corrected Schwarzschild black hole surrounded by quintessence has 
been studied recently in Ref.\ \cite{44-3}.\\[2pt]
%
\indent
As indicated earlier, the emission of GWs is an interesting phenomenon related 
to black holes, which can convey important information regarding black 
holes \cite{21,45,46,48}. QNMs represent the 
complex frequencies of oscillations associated with the emission of GWs from 
perturbed massive objects in the Universe \cite{26,27,42,45,46,48,48-1,48-2}. 
The real part of it gives the amplitude and the imaginary part represents 
the damping associated with the GWs. QNMs have been studied extensively 
in the literature in different scenarios like charged, spinning or simple 
stationary black holes with different types of surrounding fields implementing 
different theories of gravity (MTGs) apart from GR \cite{26,27,42,45,46,48,48-1,48-2}.\\[2pt] 
\indent
The greybody factor, which specifies the transmission behaviour of a black
hole is an important quantum property of black holes. This factor has been
studied extensively in different gravity theories 
\cite{27,48-3,49,50,51,53,54}. The authors of Ref.\ \cite{48-3} studied the 
Bardeen de Sitter (dS) black holes for scalar perturbation and calculated the 
greybody factors for the black holes. In Ref.\ \cite{49}, the authors have 
examined electromagnetic and gravitational perturbations of the Bardeen 
dS black hole and calculated the QNMs along with the 
greybody factor. They also examined the total absorption cross-section for 
the metric. Konoplya et.\ al.\ \cite{50} presented the recipe for computing 
the QNMs and greybody factors for black holes using the higher order 
WKB approximation method and discussed some issues as well as advantages of 
this method. The greybody factors have been studied for massive scalar fields 
in dRGT gravity for AdS and dS cases in Ref.\ \cite{51}. Authors in Ref.\ 
\cite{27} studied the \ Reissner-Nordstr\"om (RN)-type AdS/dS black holes 
with the surrounding quintessence field. In Ref.\ 
\cite{53}, the authors studied the greybody factors and Hawking radiation of 
black holes considering 4D Einstein-Gauss-Bonnet gravity. In another work, the 
authors studied the QNMs and greybody factors of black holes in 
symmergent gravity \cite{54}.\\[2pt]
\indent
In this present work, we modify the energy-momentum conservation condition of 
GR making the covariant derivative of the energy-momentum tensor proportional 
to the derivative of $R$ and $T$, where $T$ is the trace of the energy-momentum 
tensor. This model is inspired by Ref.\ \cite{40} where the authors have 
studied the effect of such a modification in Rastall theory on compact objects. 
With this modified Rastall theory we intend to study the behaviour of the 
charged black hole, i.e.\ RN black hole solution surrounded by 
a Maxwell field and a cloud of strings, specifically its QNMs, 
thermodynamics temperature and the greybody factor. Motivated by
previous studies as mentioned earlier we choose the source sustaining the 
black hole solution, i.e.\ a solution surrounded by a Maxwell field and a 
cloud of strings. The idea of black holes surrounded by a Maxwell field comes 
naturally as a black hole has a high probability of interaction with its 
surrounding environment by means of phenomena such as accretion. It is a 
physical possibility that a black hole can be charged. The strings field is 
motivated by string theory which proposes strings as the most fundamental unit 
of matter. This has been implemented in literature \cite{26} and following 
this, we implement the same in the black holes' environment in our study in 
the sense that the surrounding mass of an interacting black hole may be in 
the form of clouds of strings due to the extreme nature of black holes' 
immediate spacetime. The black hole solution obtained by us is unique and has 
scope for studying black hole shadows, accretion disk, gravitational lensing 
along with other properties.\\[2pt]
\indent  
This paper is organised as follows. In Section \ref{sec2}, we discuss the 
Rastall theory and the modification imposed. We also solve the field equations 
after considering a charged background together with string clouds. In 
Section \ref{sec3}, we give a brief account of the QNMs of oscillations of the 
black hole and compute the complex frequencies for the black hole in the 
particular setup. In Section \ref{sec4}, we compute the thermodynamic 
temperature associated with the black hole for various values of the model 
parameters and analyse them. In Section \ref{sec5}, we study the greybody 
factors and absorption coefficient associated with the black hole. Finally, in 
section \ref{sec6}, we present the concluding remarks and future directions.

\section{Field equations of modified rastall theory}\label{sec2}
GR demands that the covariant derivative of the energy-momentum tensor should 
vanish, that is $\nabla _{\nu}T^{\mu\nu} = 0.$ Rastall modified this 
conservation condition, generalising it to the form:
\begin{equation}
\nabla _{\nu}T^{\mu\nu}=\lambda \nabla^{\mu} R,
\label{a1}
\end{equation}
where $\lambda$ represents the Rastall parameter. This Rastall form of 
modification of the conservation condition is based on the fact that 
variation of energy-momentum of spacetime should depend on the corresponding
variation of the curvature of the spacetime. Nevertheless, it is also possible
that the pattern of variation of the energy-momentum of spacetime depends on 
the energy-momentum content of spacetime in addition to its curvature 
variation. Considering this aspect in mind, here we introduce a further 
modification in the conservation condition as follows:
\begin{equation}
\nabla_{\nu} T^{\mu\nu} =  \nabla^{\mu}(\lambda R+ \alpha T),
\label{a2}
\end{equation} 
where $\alpha$ is a constant parameter associated with $T$, which measures the 
deviation of the theory from Rastall's original form. Using the above 
equations, one can deduce the field equations for the modified Rastall 
gravity as
\begin{equation}
R_{\mu\nu}-\frac{1}{2}g_{\mu\nu}R + \kappa \lambda g_{\mu\nu} R + \kappa \alpha g_{\mu\nu} T = \kappa T_{\mu\nu}.
\label{a3}
\end{equation}
Assuming $\kappa \lambda=\beta$, we can rewrite the above field equations in 
an elegant form as
\begin{equation}
G_{\mu\nu}+\beta g_{\mu\nu}R + \kappa \alpha g_{\mu\nu} T = \kappa T_{\mu\nu}.
\label{a4}
\end{equation}
The trace of the equation \eqref{a4} gives
\begin{equation}
R=\frac{\kappa T(1-4\alpha)}{4\beta-1}.
\label{a5}
\end{equation}
For our work, we will stick to the spherically symmetric black hole metric 
ansatz \cite{26},
\begin{equation}
ds^2 = -f(r)\, dt^2 + \frac{dr^2}{f(r)}+r^2 d\Omega^2,
\label{a6}
\end{equation}
where $d\Omega^2=d\theta^2 +\sin^2 \theta\, d\phi^2$.
Defining the Rastall tensor $\Theta_{\mu\nu}=G_{\mu\nu}+\beta g_{\mu\nu}R +
\kappa \alpha g_{\mu\nu} T$, the following non-vanishing components of the modified 
field equations \eqref{a4} can be obtained as
\begin{align}
\Theta^0_0&=\frac{1}{r^2}\Big(r f'(r) + f(r) -1\Big) + \beta R + \kappa \alpha T,
\label{a7}\\[5pt]
\Theta^1_1&=\frac{1}{r^2}\Big(r f'(r) + f(r) -1\Big) + \beta R + \kappa \alpha T,
\label{a8}\\[5pt]
\Theta^2_2&=\frac{1}{r^2}\Big(r f'(r) + \frac{1}{2}r^2 f''(r)\Big) + \beta R + \kappa \alpha T,
\label{a9}\\[5pt]
\Theta^3_3&=\frac{1}{r^2}\Big(r f'(r) + \frac{1}{2}r^2 f''(r)\Big) + \beta R + \kappa \alpha T.
\label{a10}
\end{align}
Here, prime denotes derivative with respect to $r$. The expression of the 
Ricci scalar is obtained as
\begin{equation}
R=-\frac{1}{r^2}\Big(r^2 f''(r) +4r f'(r) +2f(r)-2\Big).
\label{a11}
\end{equation}

We assume that the spherically symmetric spacetime surrounding the
black hole is characterized by the presence of an electric charge field and
a string's cloud field. Hence, the total energy-momentum tensor of the 
considered spacetime is defined as
\begin{equation}
T^{\mu}_{\nu}=E^{\mu}_{\nu}+ \mathcal{T}_{\nu}^{\mu}.
\label{a12}
\end{equation}
In this relation $E^{\mu}_{\nu}$ is the Maxwell tensor having the form:
\begin{equation}
E^{\mu}_{\nu}=\frac{q^2}{\kappa\, r^4}
\begin{pmatrix}
-1 & 0 & 0 & 0\\
0 & -1 & 0 & 0\\
0 & 0 & 1 & 0\\
0 & 0 & 0 & 1\\
\end{pmatrix},
\label{a13}
\end{equation}
where $q$ denotes the black hole charge parameter. The other term 
$\mathcal{T}^{\mu}_{\nu}$ represents the surrounding string's cloud field 
which in simplified form can be written as \cite{26}
\begin{equation}
\mathcal{T}^{\mu}_{\nu}=
\begin{pmatrix}
\rho_c(r) & 0 & 0 & 0\\
0 & \rho_c(r) & 0 & 0\\
0 & 0 & 0 & 0\\
0 & 0 & 0 & 0\\
\end{pmatrix},
\label{a14}
\end{equation}
where $\rho_c$ is the string's cloud density parameter. To derive 
the explicit form of this parameter, we make use of the equations \eqref{a2}, 
\eqref{a5} and \eqref{a14} to get the following differential equation:
\begin{equation}
\frac{\partial \rho_c}{\partial r}+\frac{2 \rho_c}{r}= \bigg[\frac{2\beta(1-4\alpha)}{4\beta-1}+2\alpha\bigg]\frac{\partial \rho_c}{\partial r}.
\label{a15}
\end{equation}
The above equation can be solved and the following solution for 
$\rho_c$ is obtained:
\begin{equation}
\rho_c (r)=b\, r^{\frac{2(1-4\beta)}{2\alpha (1+8\beta)-2\beta-1}},
\label{a16}
\end{equation}
where $b$ is the constant of integration, which is associated with the string's 
density parameter. We impose the condition on $b$ that $b\geq 0$ to respect 
the weak energy condition. Moreover, one interesting point to note from 
equation \eqref{a16} is that for the physical consistency, we should have 
$2\alpha (1+8\beta)-2\beta-1 > 0$, otherwise its equality to zero will lead to 
divergence of string density. This condition leads to setting a constraint on 
the possible values of $\alpha$ for a given value of $\beta$. For instance, if
we choose $\beta=0.01$, we see that the value of $\alpha$ should be $> 0.47$. 
Finally, we are in a situation to write the Rastall field equations in a 
complete form as in the following:
\begin{align}
\frac{1}{r^2}\Big(r f'(r) & +f(r)-1\Big) - \frac{\beta }{r^2}\Big(r^2 f''(r) +4r f'(r) +2f(r)-2\Big) +\kappa \alpha T=\kappa \rho_c -\frac{q^2}{r^4},
\label{a17}\\[8pt]
\frac{1}{r^2}\Big(r f'(r) & +\frac{1}{2}r^2 f''(r)\Big) -  \frac{\beta}{r^2}\Big(r^2 f''(r) +4r f'(r) +2f(r)-2\Big) +\kappa \alpha T=\frac{q^2}{r^4}.
\label{a18}
\end{align}
Solving equations \eqref{a17} and \eqref{a18}, the general solution for the 
metric function of the black hole is found as
\begin{equation}
f(r)=1-\frac{2 M}{r}+\frac{q^2}{r^2}+\frac{a (-2 \alpha  (8 \beta +1)+2 \beta +1)^2 r^{\frac{(4 \alpha -1) (8 \beta +1)}{2 \alpha  (8 \beta +1)-2 \beta -1}}}{(4 \beta -1) (6 \alpha  (8 \beta +1)-14 \beta -1)},
\label{a19}
\end{equation}
where $a=\kappa b$ is the string parameter as it is directly associated with 
the density of the string field. This string parameter $a$ is constrained by 
the weak energy condition as $a\ge 0$. The black hole metric function 
\eqref{a19} is the RN black hole solution surrounded by a cloud of strings in 
the modified Rastall theory. It is to be noted at this point that a
feasible black hole solution should satisfy the energy conditions, especially
the weak energy condition (WEC). Thus we need to check the consistency of 
solution \eqref{a19} with the WEC. The criteria for satisfaction of WEC are as 
follows \cite{weak}:

\begin{equation}
\frac{2}{r} \frac{d m(r)}{dr}\geq \frac{d^2 m(r)}{dr^2},
\label{a19-1}
\end{equation}
\begin{equation}
\frac{1}{r^2}\frac{dm(r)}{dr}\geq 0.
\label{a19-2}
\end{equation}
Here $m(r)$ represents the mass function that can be extracted 
from the metric solution and has the following form.

\begin{equation}
m(r)= M-\frac{q^2}{2r}+\frac{a r^{\frac{2(\alpha-3\beta+8\alpha \beta)}{2\alpha(1+8\beta)-2\beta-1}} (1+2\beta-2\alpha(1+8\beta))^2}{2(1-4\beta)(1+14\beta-6\alpha(1+8\beta))}.
\label{a19-2}
\end{equation}

\begin{figure}
\centerline{
\includegraphics[scale=0.4]{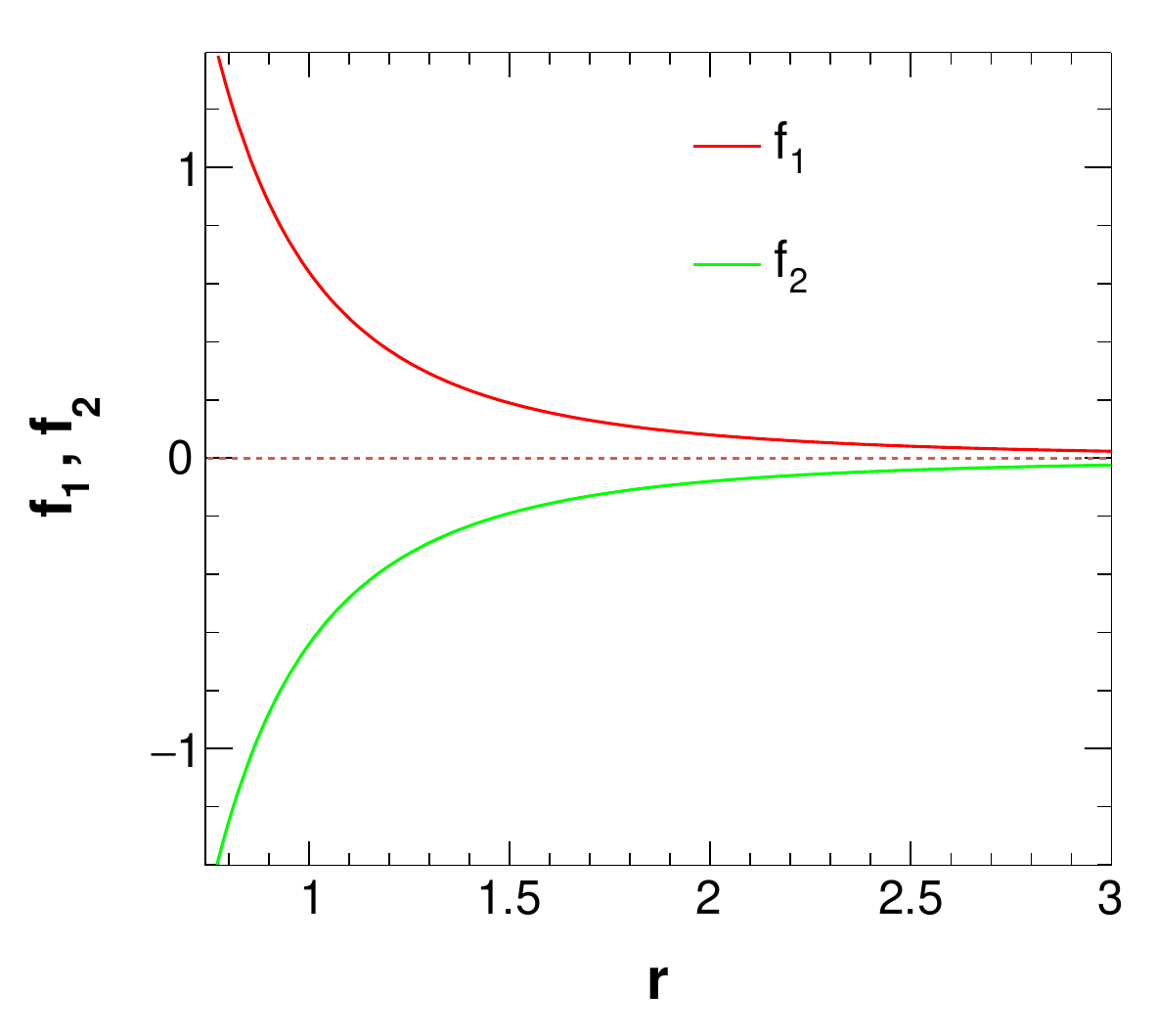}\hspace{0.5cm}
\includegraphics[scale=0.4]{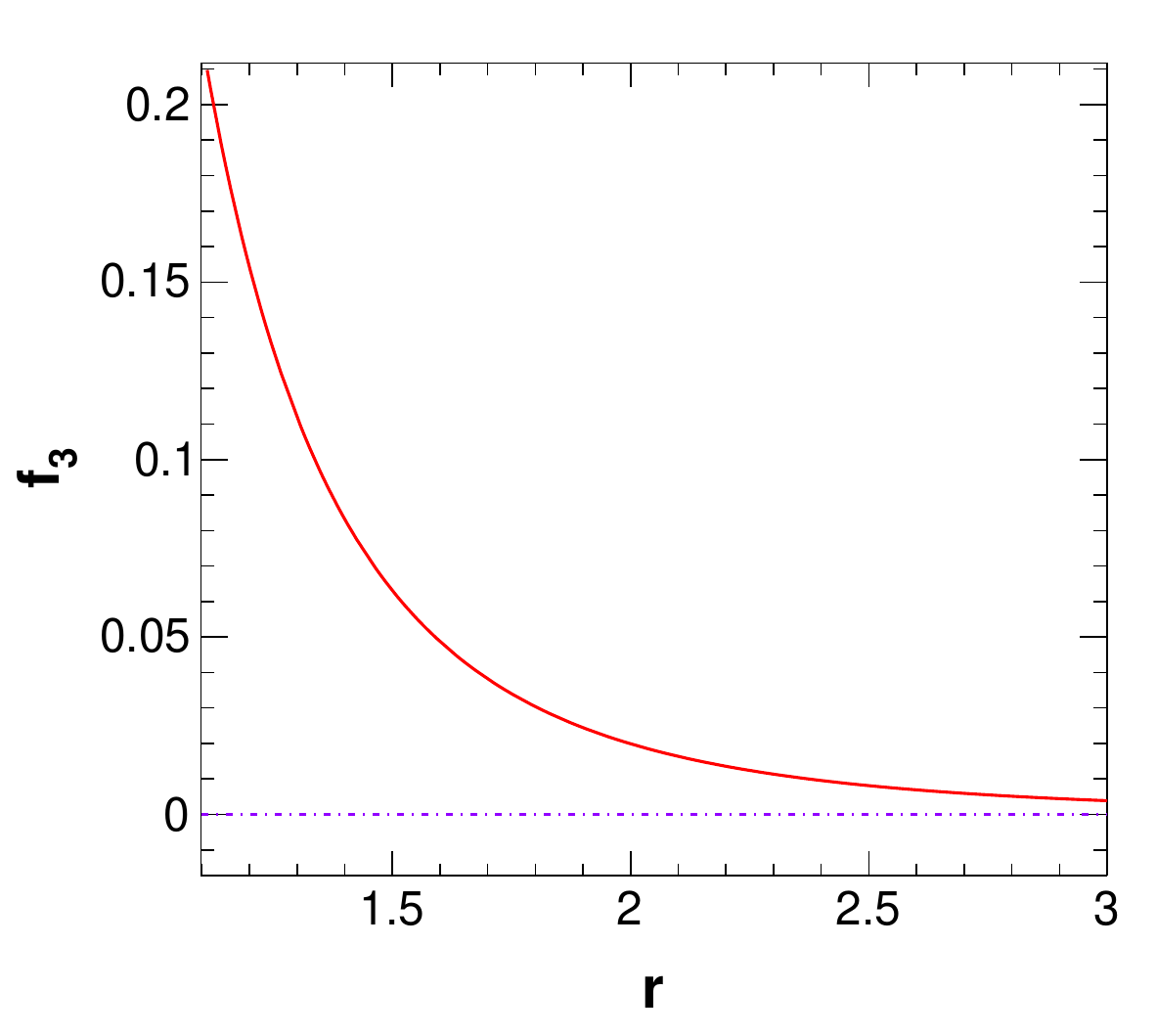}
}
\vspace{-0.2cm}
\caption{Weak energy condition for the black hole solution \eqref{a19}. 
Functions $f_1$, $f_2$ and $f_3$ have been defined in the text. Parameter 
values $a=0.001$, $\alpha=0.8$, $\beta=0.01$ and $q=0.80$ have been used.}
\label{02b}
\end{figure}

Calling these three functions as $f_1 = 2d m(r)/rdr$,
$f_2=d^2 m(r)/dr^2$ and $f_3=dm(r)/r^2dr$ respectively, we have shown in 
Figure \ref{02b} the obeyance of the WEC by the solution \eqref{a19} for the 
parameter values we mostly use in this work. It should also be noted that 
violation of WEC has been observed for higher values of parameters.

\begin{figure}[h!]
\centerline{
\includegraphics[scale=0.7]{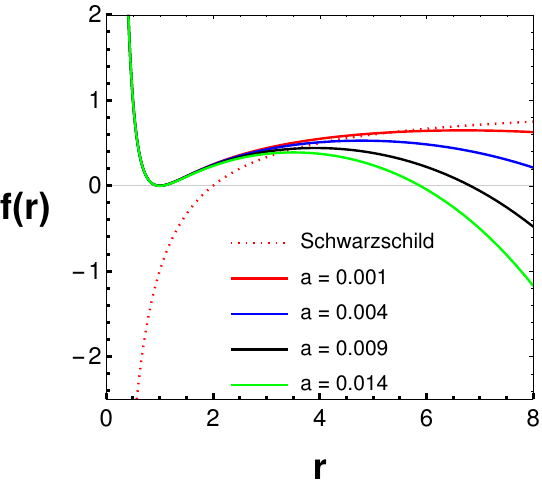}\hspace{1.0cm}
\includegraphics[scale=0.7]{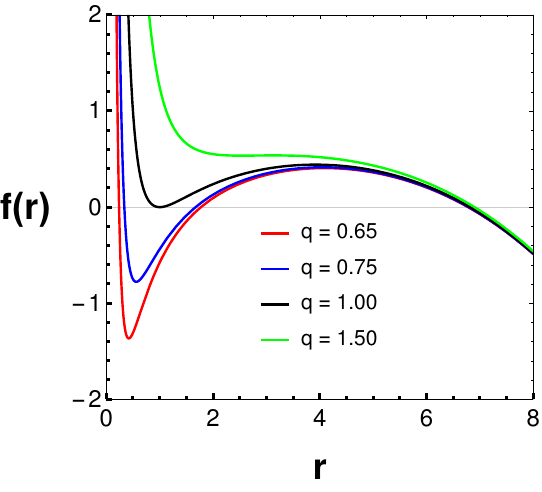}}\vspace{0.5cm}
\centerline{
\includegraphics[scale=0.7]{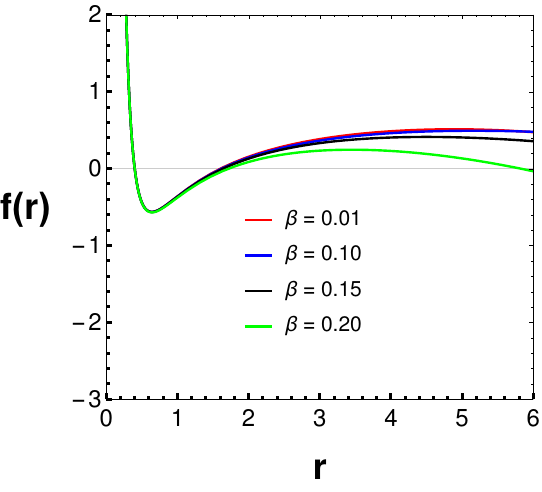}\hspace{1.0cm}
\includegraphics[scale=0.7]{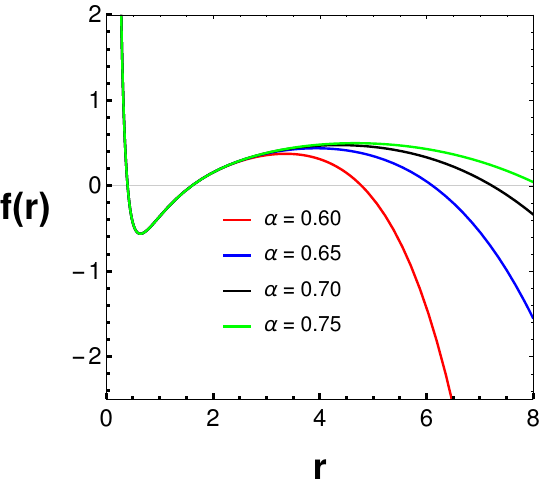}}
\vspace{-0.2cm}
\caption{Behaviour of the metric function of the black hole with respect to
distance $r$ for different values of the model parameters. For the first plot, 
we use $\beta=0.01$, $q=1$, $\alpha =0.80$, for the second one, we use 
$\beta=0.01$, $a=0.009$, $\alpha=0.80$, for the third plot, $a=0.004$, 
$\alpha=0.80$ and $q=0.8$ are used and for the fourth one, we use $a=0.004$, 
$\beta=0.01$ and $q=0.80$. Here we set $M=1$ and the same will be followed for 
all remaining plots.}
\label{01}
\end{figure}

Returning to equation \eqref{a19}, in the limit $a$ and $\beta$ going 
to zero, we recover the RN solution and also for $q=0$, we recover the 
Schwarzschild black hole solution. We can also recover the well-known 
RN-AdS solution from the metric solution by properly substituting values of 
the constants. This shows that the metric solution obtained here is general 
and encompasses many well-known black hole solutions. The behaviour of the 
metric function \eqref{a19} is shown in Figure \ref{01} with various values 
of the model parameters. Here, we can see that there are three 
horizons for the black hole metric. It is observed that the outermost horizon 
is impacted mostly by string parameter $a$ and hence it is called the string 
horizon. The first plot 
shows the metric function variation for different values of the string 
parameter $a$. It is seen from the curves that variation in the values of $a$ 
does not affect the single inner horizon of the black hole but there is a 
visible impact on the outer horizon, which we may refer to as the string 
horizon. With the increase in $a$, the string horizon gradually decreases as 
shown in the plot. The second plot shows the metric function variation for 
various values of charge parameter $q$. Here, it is seen 
that $q$ has negligible 
influence on the string horizon but mainly influences the two inner horizons. 
As $q$ increases, the dip of the metric curve decreases and finally, we get a 
critical value of $q=1$ beyond which there is no inner horizon as can be seen 
from the plot. The third plot shows the behaviour of the metric function for 
various values of the Rastall parameter $\beta$. It is clear that $\beta$ has 
negligible influence on the inner horizons but mainly impacts the outer string 
horizon. As can be seen, with higher values of $\beta$, the outer horizon 
radius decreases gradually. Also, we plot the metric function for different 
values of $\alpha$. As can be seen from the fourth plot, with an increase in 
$\alpha$, the outer horizon radius increases. Further, it is to be noted that 
as seen from the first plot of the figure, the metric function behaves very 
differently than the Schwarzschild one except in the overlapping 
region of the horizon radius.   

\begin{figure}[h!]
\centerline{
\includegraphics[scale=0.28]{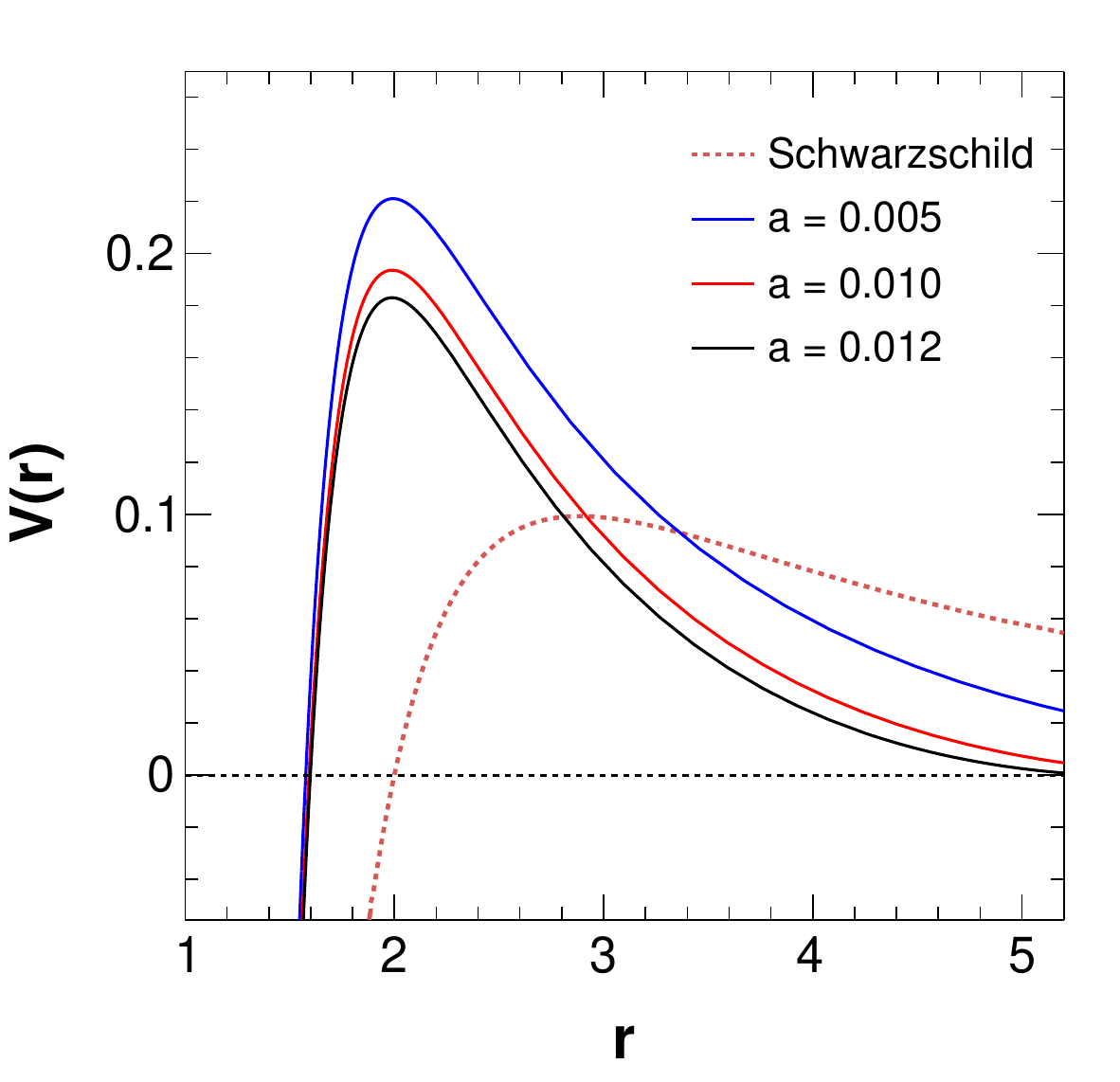}\hspace{0.3cm}
\includegraphics[scale=0.28]{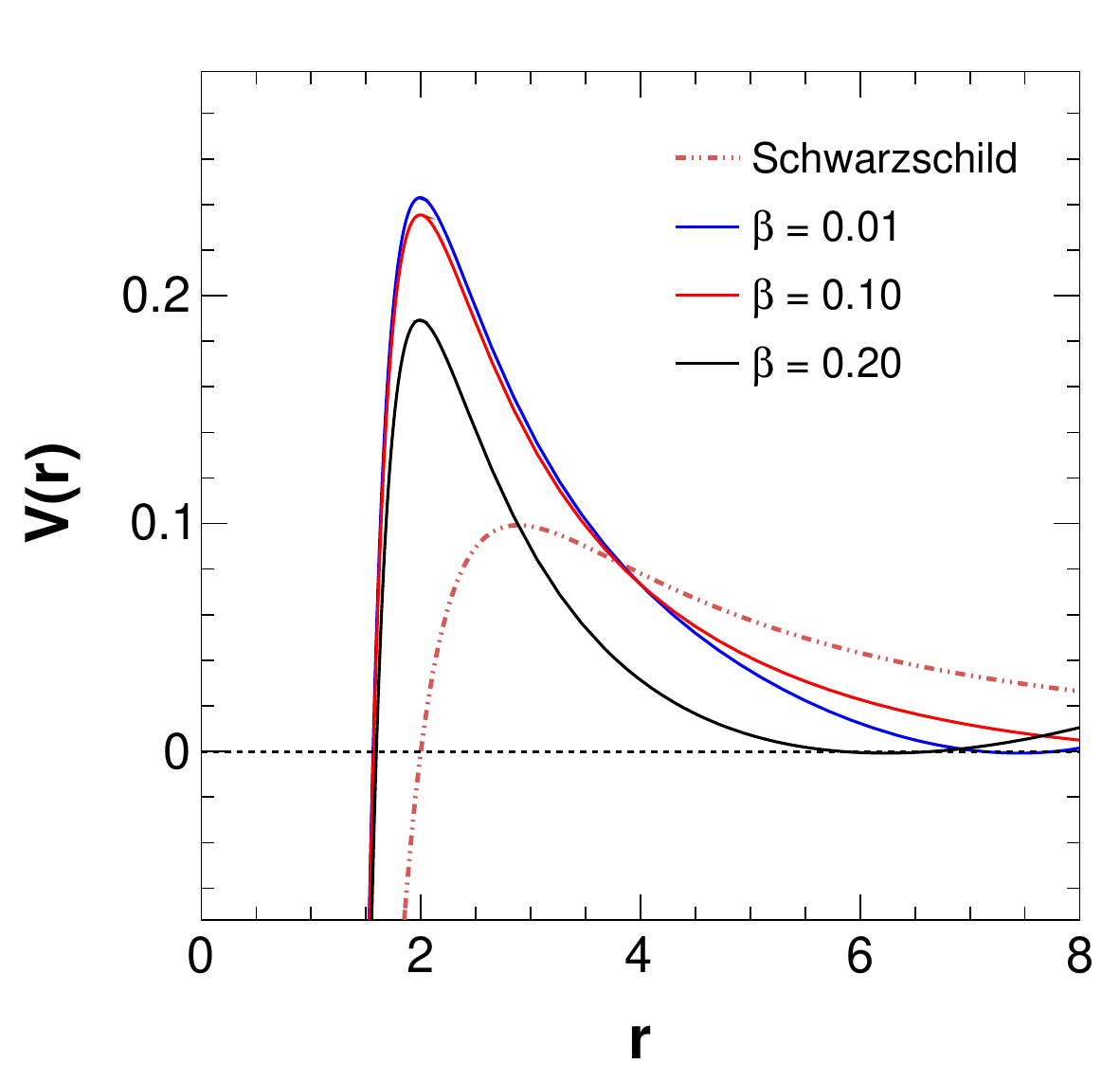}\hspace{0.3cm}
\includegraphics[scale=0.28]{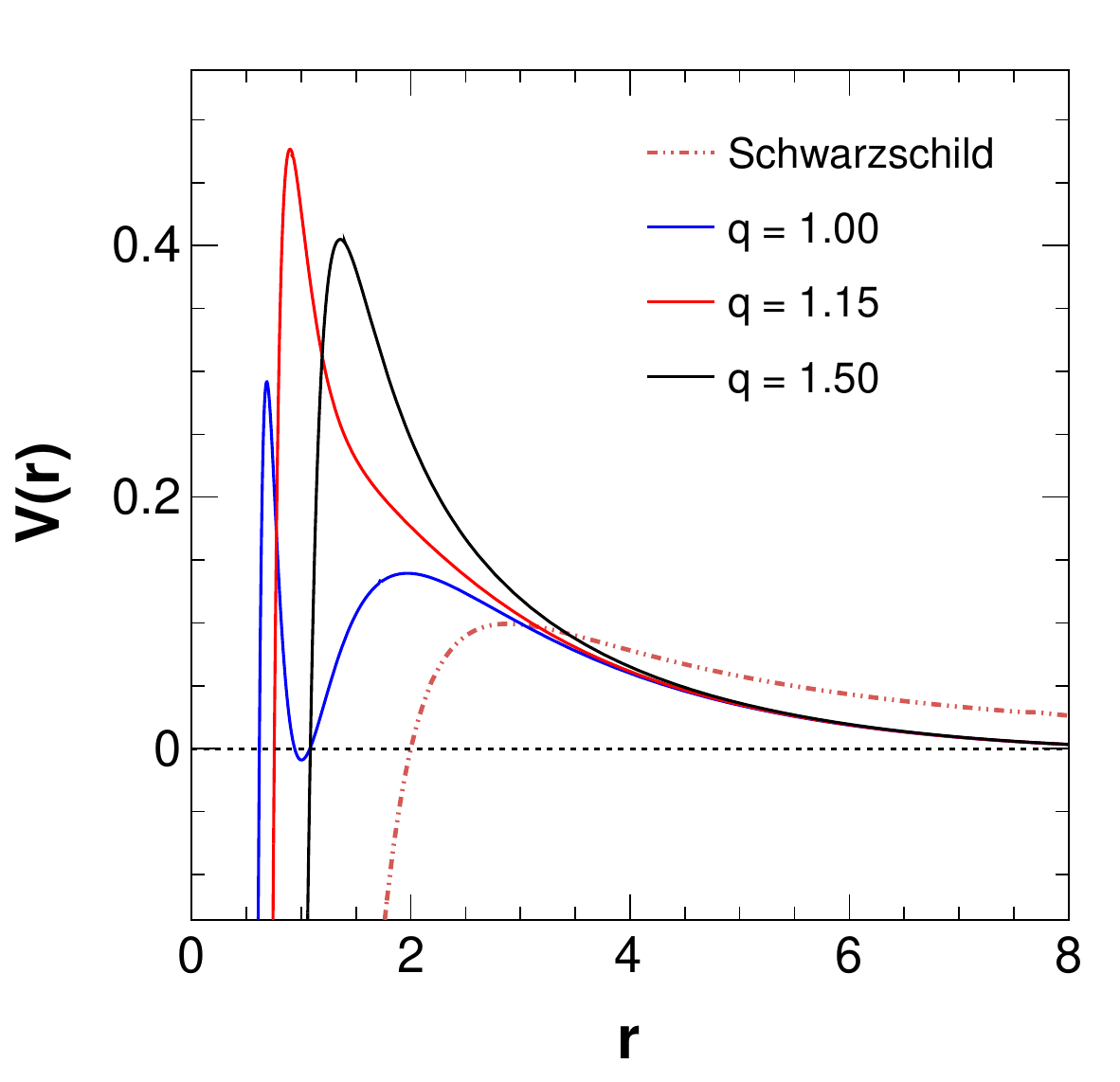}\vspace{0.2cm}
}
\centerline{
\includegraphics[scale=0.28]{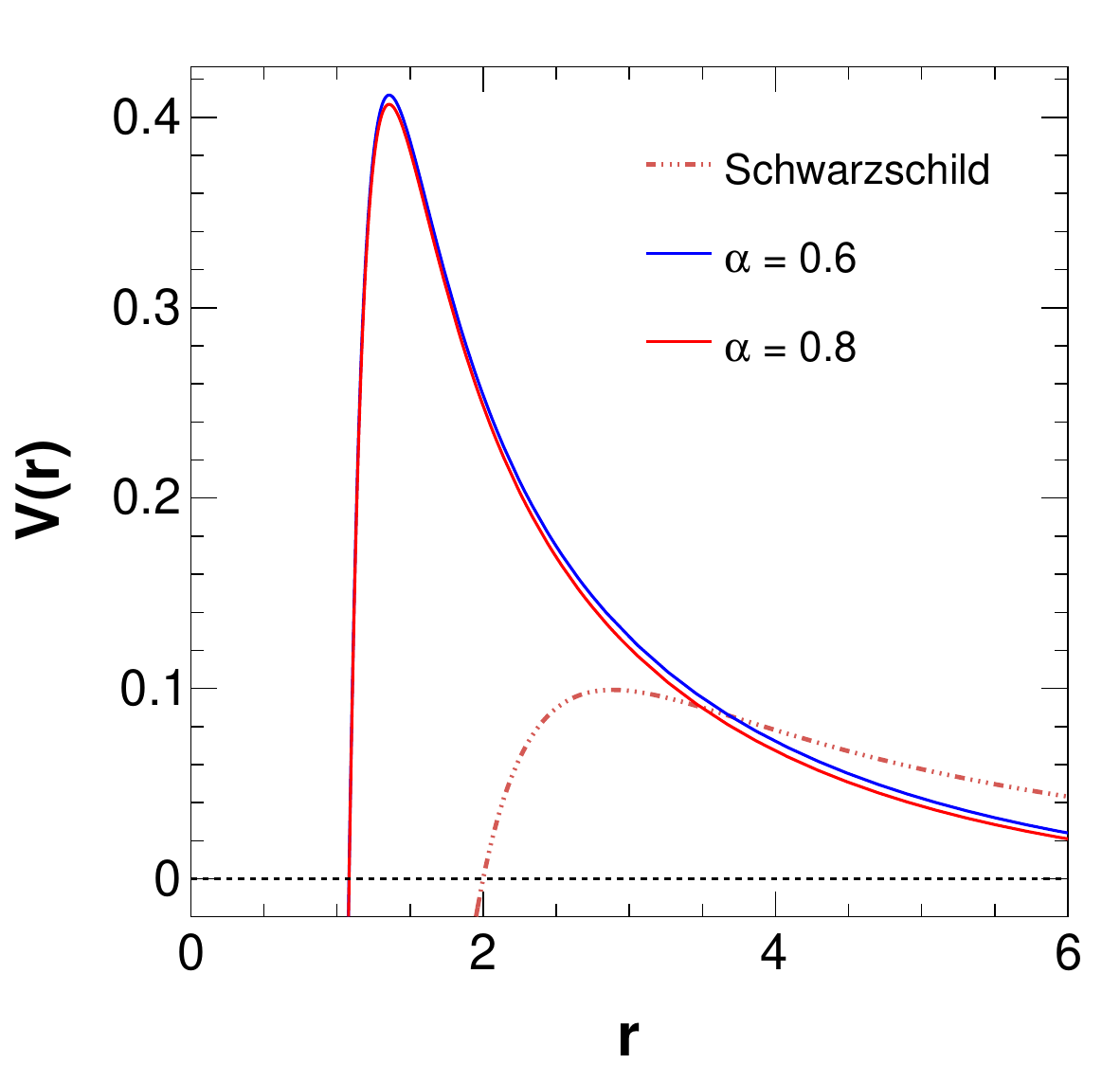}\hspace{0.3cm}
\includegraphics[scale=0.28]{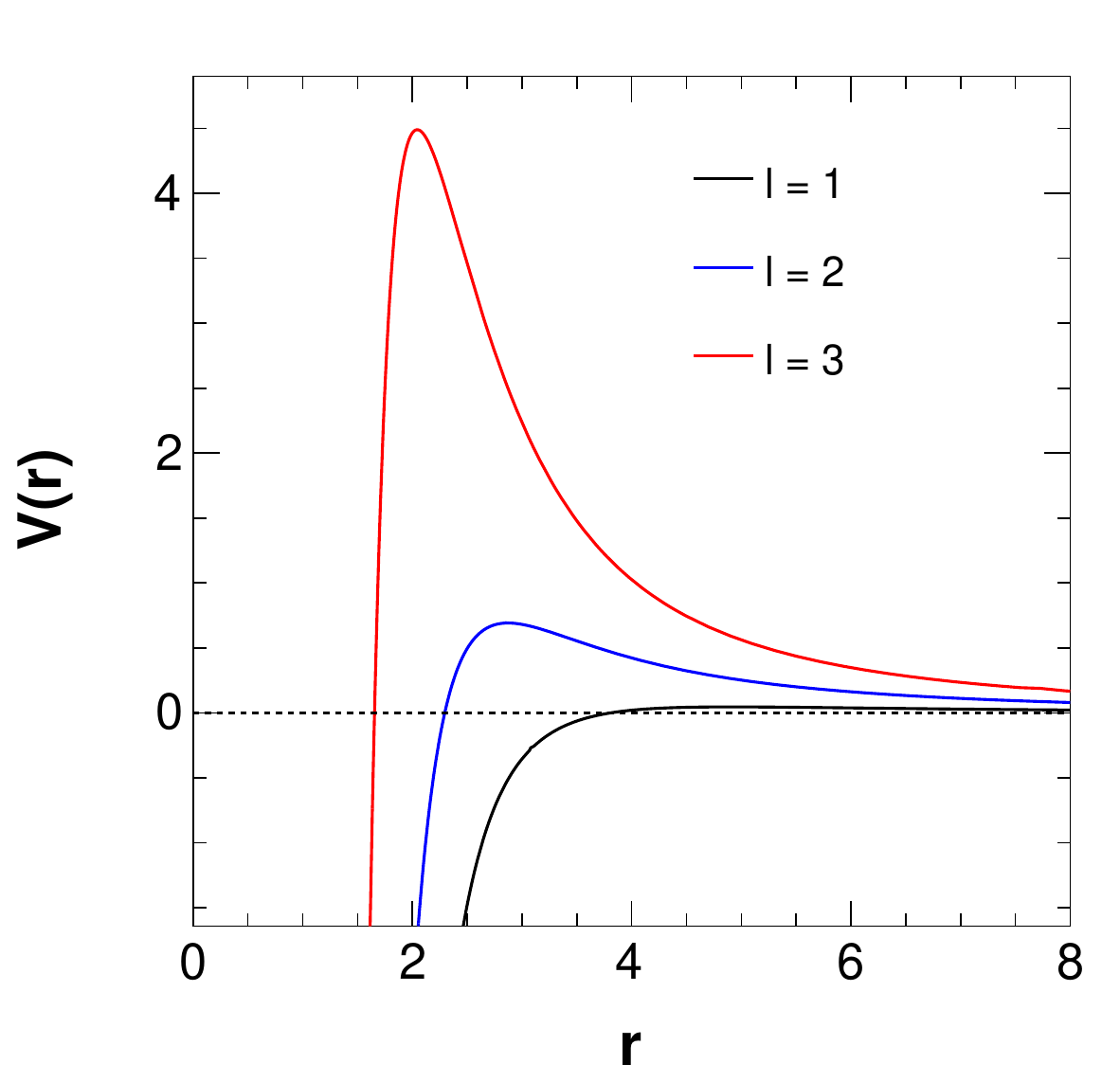}
}
\vspace{-0.2cm}
\caption{Behaviour of the potential \eqref{b5} for different values of the 
model parameters. For the top left plot, we use $\beta=0.15$, $q=2$, 
$\alpha=0.80$, for the top middle plot, we use $a=0.005$, $q=2$, 
$\alpha=0.75$, for the top right plot, we use $a=0.005$, $\beta=0.12$, 
$\alpha=0.75$, and for the bottom left plot, we use $a=0.005$, $\beta=0.1$, 
$q=1.5$. In all these three plots we consider $l=1$. In the bottom right plot 
$a=0.001$, $\beta=0.1$, $q=4.25$ and $\alpha = 0.8$ are used.}
\label{02a}
\end{figure}
Now we focus on the determination of the black hole's QNMs of oscillations for 
which we shall use the 6th order WKB approximation method. This method is the 
most widely used and trusted method of calculating QMNs and provides a 
mechanism for error estimation as well as higher order calculations for 
better accuracy. It matches the results of other analytical and numerical 
methods like the asymptotic iteration method, continued fraction method and 
time domain analysis method \cite{27,45,49,aim}. 

\section{WKB approximation method for calculating QNMs}\label{sec3}
The most important step for computing the QNMs of the black hole defined 
by the metric function \eqref{a19} with the scalar perturbation using the WKB
approximation method is to obtain the potential associated with the black hole 
metric. For which we are to perturb the black hole spacetime with a probe that 
minimally couples with a scalar field $\Phi$ described by the equation of 
motion \cite{42},
\begin{equation}
\frac{1}{\sqrt{-g}} \partial_{a}(\sqrt{-g} g^{a b}\partial_b)\Phi=0.
\label{b1}
\end{equation}
Here we consider a massless scalar field so that the right hand side of the 
above equation reduces to zero. In this setup, it is convenient to express 
field $\Phi$ in spherical polar form as \cite{42}
\begin{equation}
\Phi(t, r, \theta, \phi)=\exp^{-i \omega t}\frac{\psi(r)}{r}\,Y_l^m (\theta, \phi), 
\label{b2}
\end{equation}
where we represent the radial part of the wave by $\psi$ and $Y_l^m$ 
represents the spherical harmonics. $\omega$ is the oscillation frequency of
the time component of the wave, which corresponds to the frequency of QNMs of 
oscillation of the black hole solution. Implementing equation \eqref{b2} in 
equation \eqref{b1}, one can obtain the following Schr\"odinger-type equation:
\begin{equation}
\frac{d^2 \psi}{dx^2} +(\omega^2 -V_l(r))\psi=0,
\label{b3}
\end{equation}
where the new variable $x$ is the well-known tortoise coordinate, defined as
\begin{equation}
x=\int\! \frac{dr}{f(r)}
\label{b4}
\end{equation}
and the effective black hole potential $V_l(r)$ for the setup is obtained from 
the usual from \cite{42}:
\begin{equation}
V_l(r)=f(r)\Big(\frac{f'(r)}{r}+\frac{l(l+1)}{r^2}\Big).
\label{b5}
\end{equation}
Here $l$ denotes the multipole number. Figure \eqref{02a} shows the behaviour
of this potential for different values of model parameters $a$, $\alpha$, 
$\beta$ and 
$q$ along with different values of the multipole number $l$. It is seen that 
for all model parameters, the potential is significantly different from that 
for the Schwarzschild case, especially in the peak region of the potential. 
Moreover, for higher $l$ values the peak of the potential is substantially 
higher than that for the smaller $l$ and it also shifts towards the smaller
horizon radius.   

In order to have physical consistency, we impose boundary conditions on the 
radial part of the wave function at the horizon and infinity as follows:
\begin{align}
\psi(x) \rightarrow \Bigg \{ \begin{array}{ll}
A e^{+i\omega x} & \text{if }x \rightarrow -\infty \\[3pt]
B e^{-i\omega x} & \text{if }x \rightarrow +\infty, \end{array}
\label{b6}
\end{align} 
where $A$ and $B$ are the integration constants. In consideration of the above 
criteria, we calculate the QNM frequencies following the Refs.\ \cite{42,45,46,48}. 
In the following figures, we plot the QNM frequencies for the black 
hole with variations in the model parameters to show their impact on the 
amplitude and damping of the QNMs. 

\begin{figure}[h!]
\includegraphics[scale=0.35]{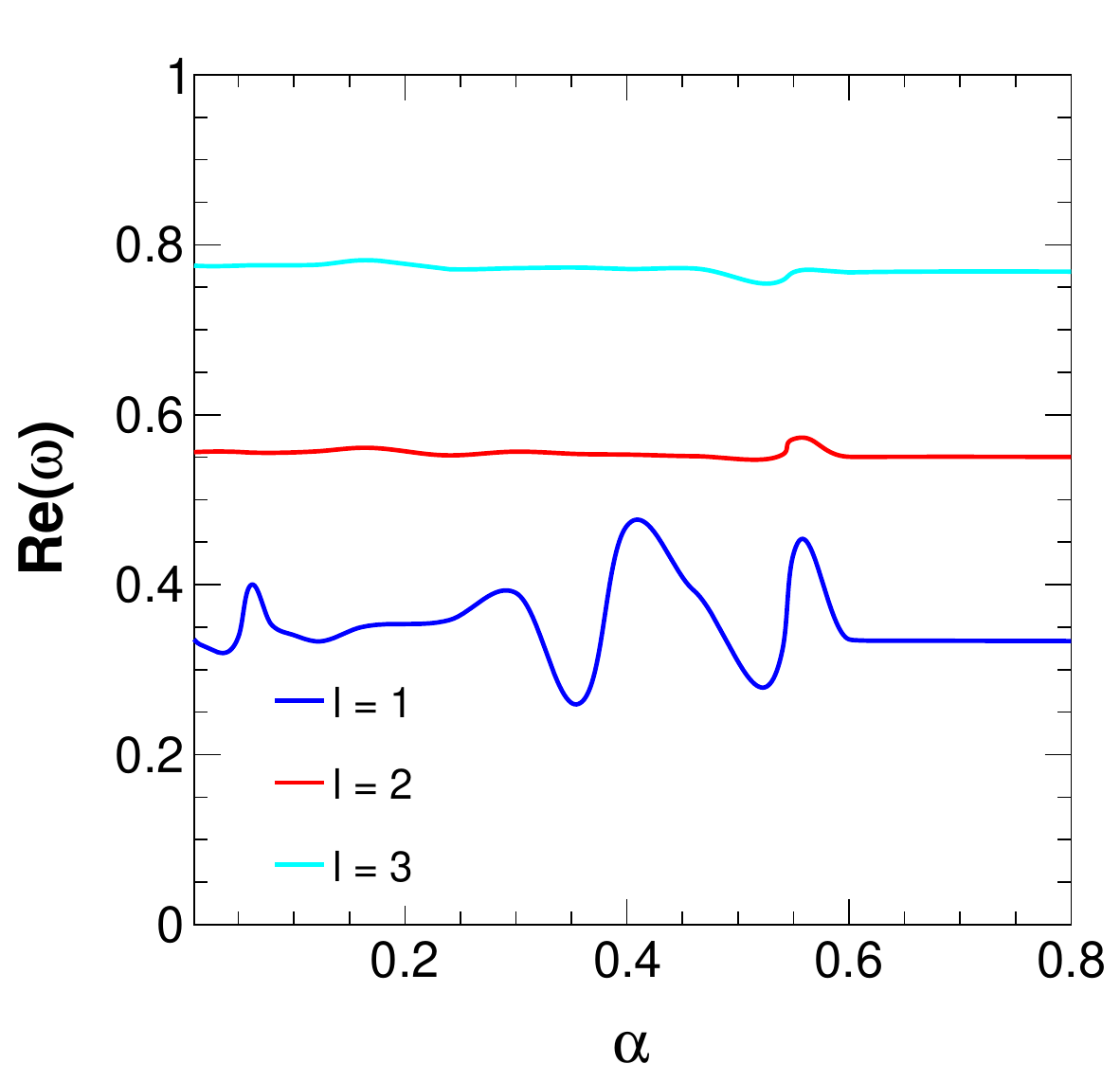}\hspace{0.5cm}
\includegraphics[scale=0.35]{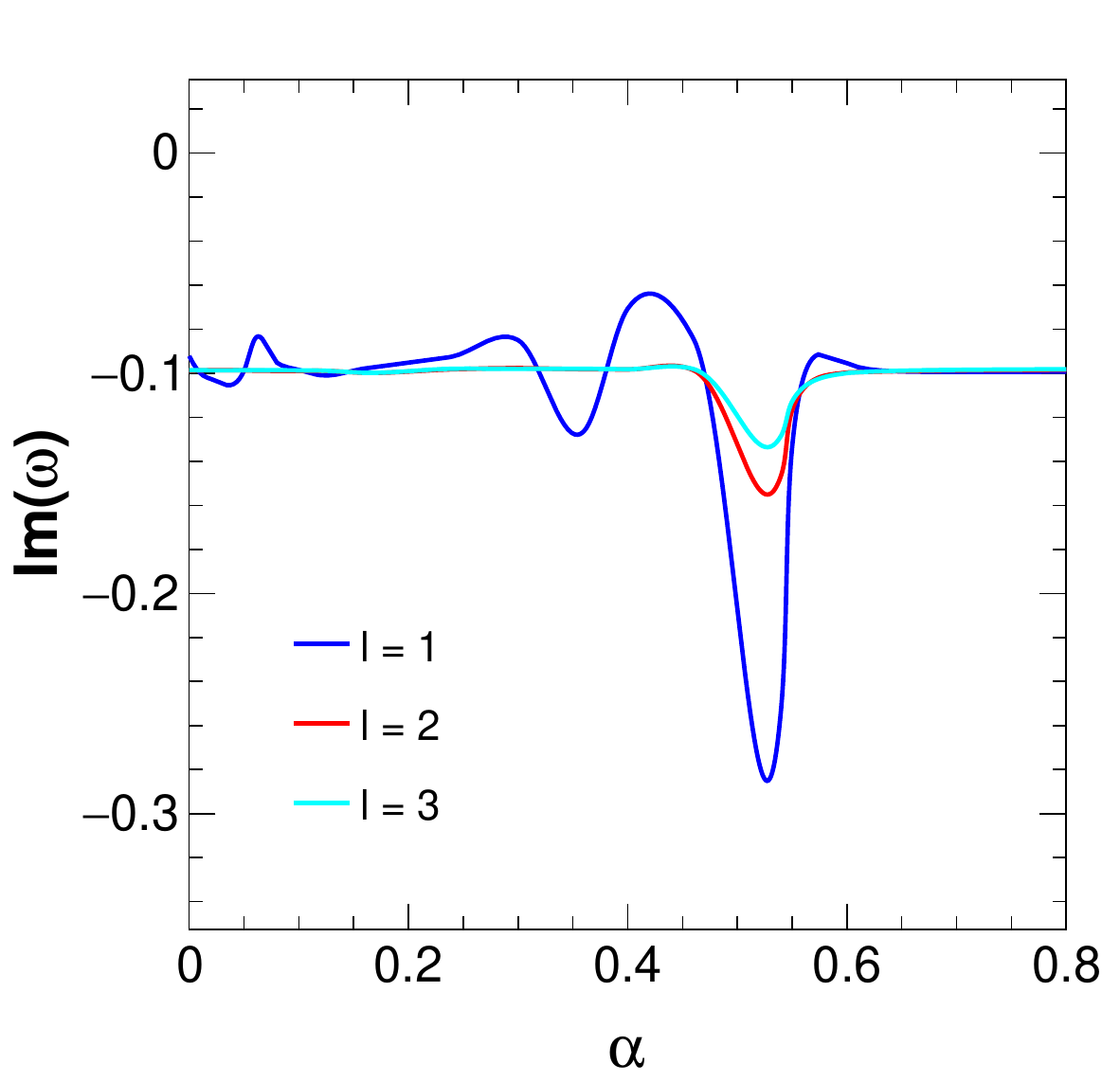}
\vspace{-0.2cm}
\caption{Variation of QNM frequencies with respect to parameter $\alpha$ for
different values of $l$. The left plot is for the amplitude part and the right 
plot is for the damping part obtained by taking $q=0.80$, $a=0.001$ and 
$\beta=0.01$.}
\label{02-1}
\end{figure}

\begin{figure}[h!]
\includegraphics[scale=0.35]{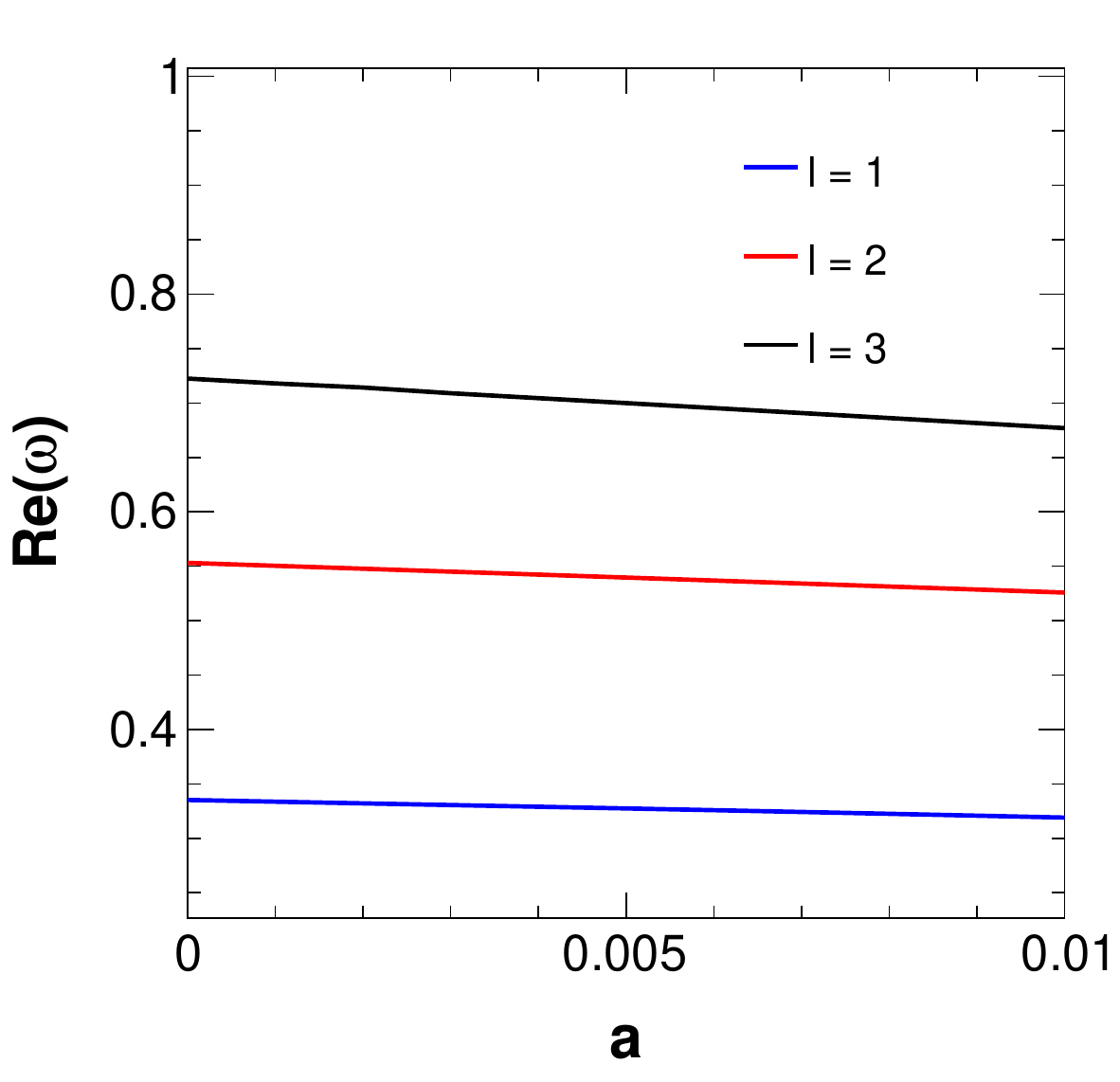}\hspace{0.5cm}
\includegraphics[scale=0.35]{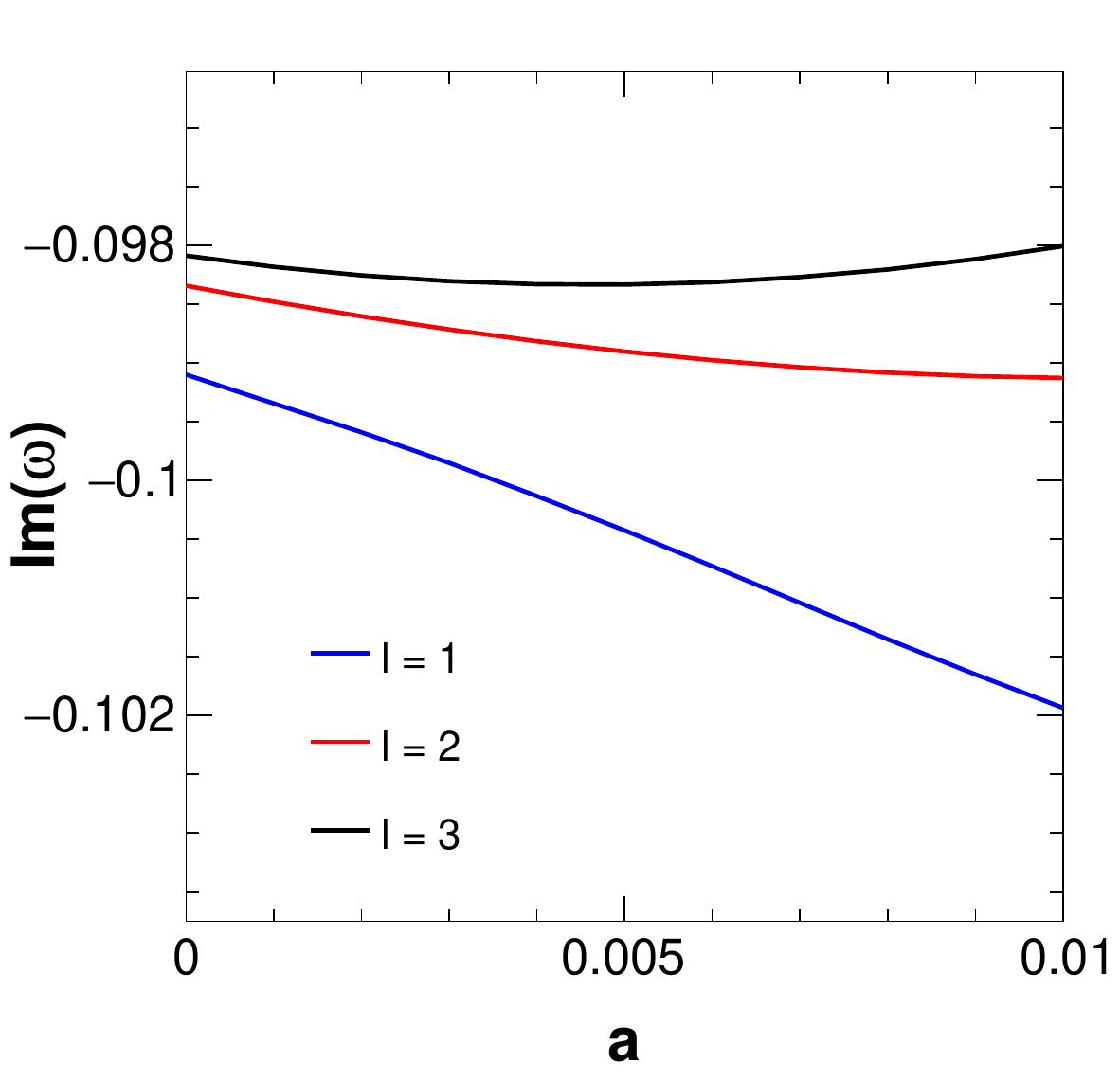}
\vspace{-0.2cm}
\caption{Variation of QNM frequencies with respect to string parameter 
$a$ for different values of $l$. The left plot is for the amplitude part and 
the right plot is for the damping part obtained by taking $q=0.8$, 
$\alpha=0.80$ and $\beta=0.01$.}
\label{02}
\end{figure}

\begin{figure}[h!]
\includegraphics[scale=0.35]{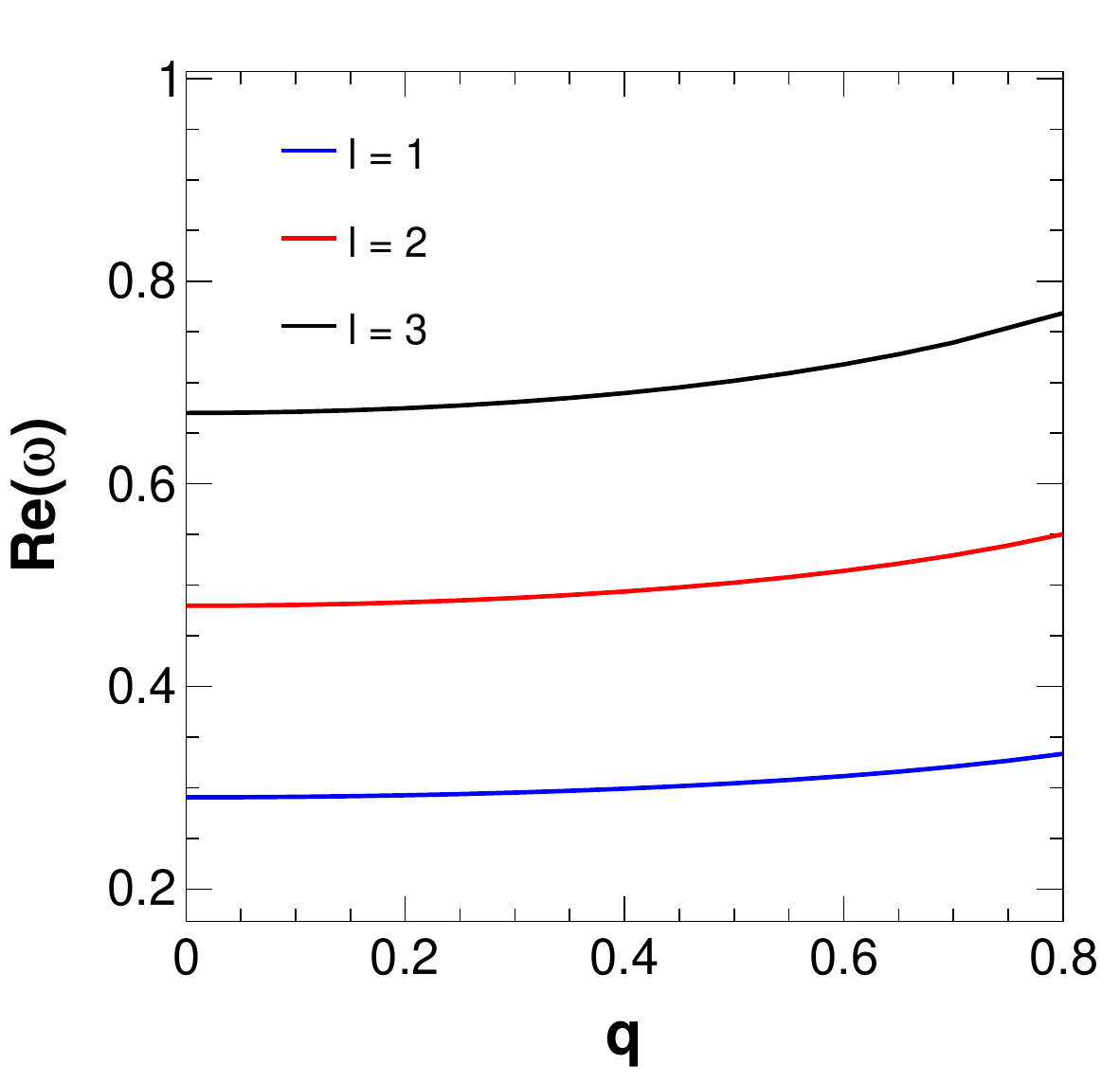}\hspace{0.5cm}
\includegraphics[scale=0.35]{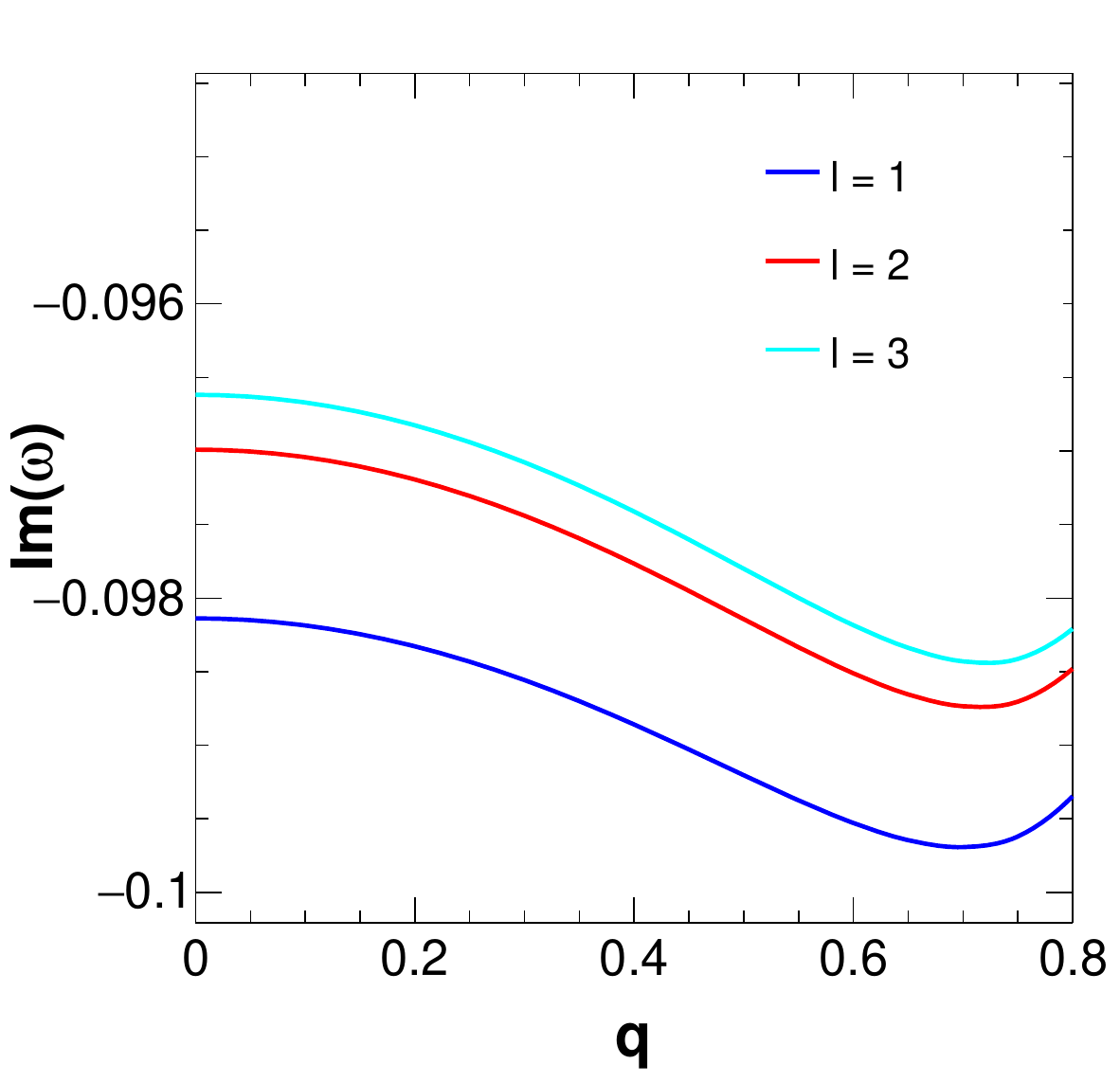}
\vspace{-0.2cm}
\caption{Variation of QNM frequencies with variation in $q$ for different 
values of $l$. The left plot shows the variation of the amplitude part and the 
right plot shows the variation of the damping part for $a=0.001$, 
$\alpha=0.80$ and $\beta=0.01$.}
\label{03}
\end{figure}

\begin{figure}[h!]
\includegraphics[scale=0.35]{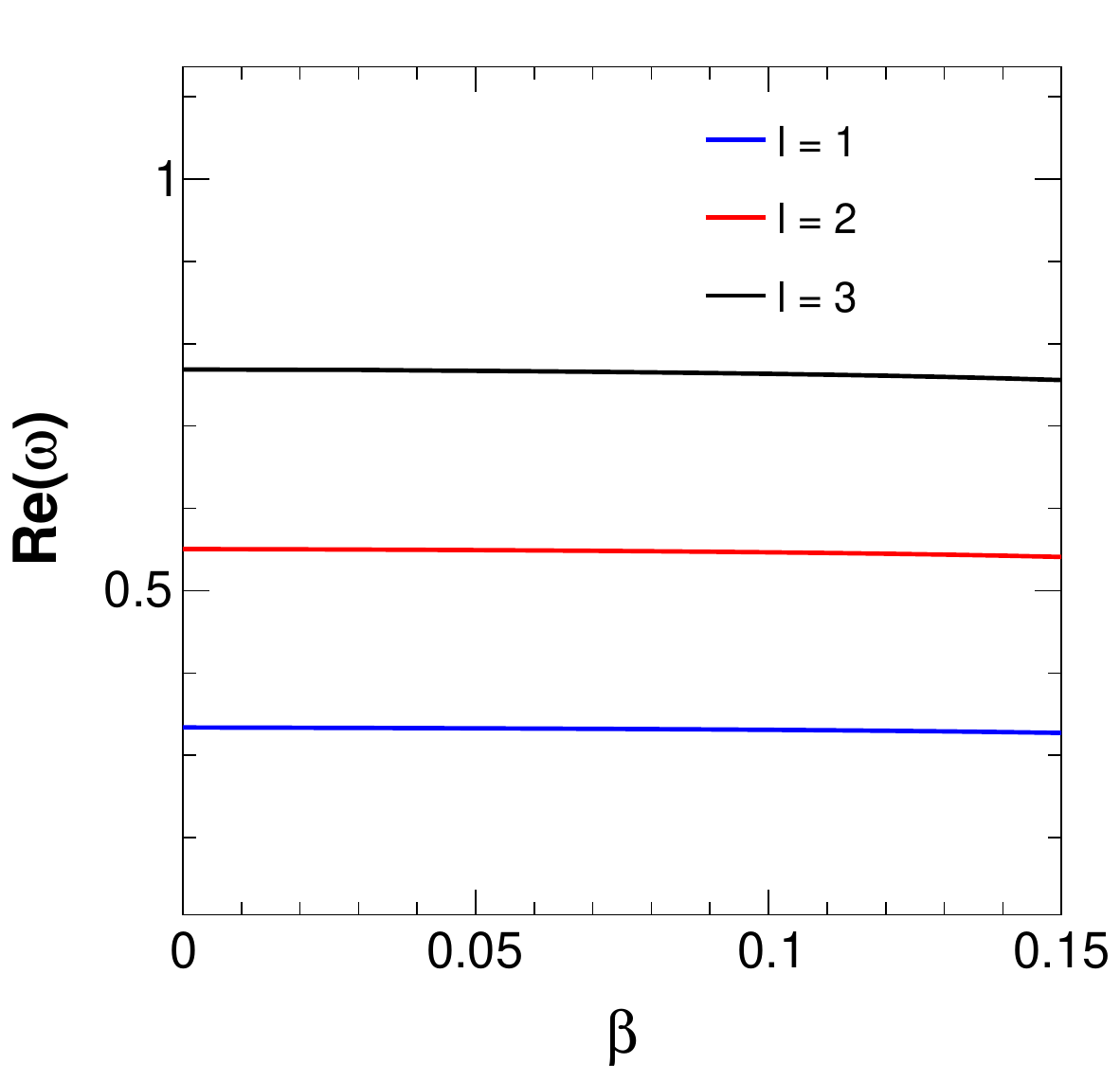}\hspace{0.5cm}
\includegraphics[scale=0.35]{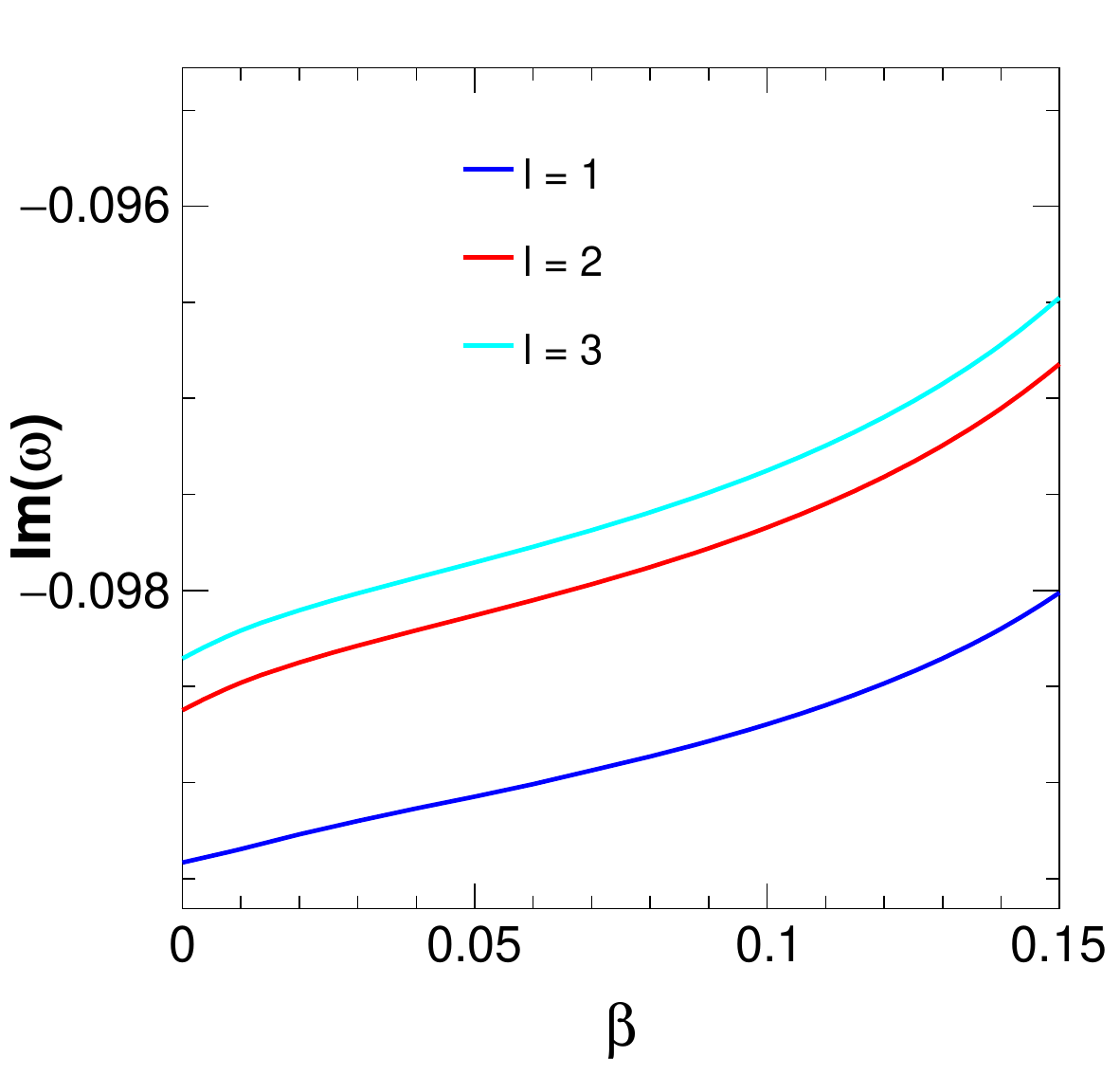}
\vspace{-0.2cm}
\caption{Variation of QNM frequencies with variation in $\beta$ for different 
values of $l$. The left plot shows the variation of the amplitude part and the 
right plot shows the variation of the damping part for $q=0.8$, $\alpha=0.80$ 
and $a=0.001$.}
\label{04}
\end{figure}
Figure \ref{02-1} shows the variation of amplitude and damping of QNMs with 
variation in the parameter $\alpha$. It is seen that there are random 
oscillatory behaviours in both the real and imaginary parts of the QNMs for 
values of $\alpha < 0.6$. For values of $\alpha > 0.6$, the QNM frequencies 
show the non-oscillatory stable behaviour. Moreover, the oscillatory behaviour
of QNMs for the said range of values of $\alpha$ reduces with the increase 
in values of $l$. This oscillatory behaviour of QNMs signifies the unstable 
nature of the black hole solution for that particular range of values of 
$\alpha$, and this also supports our earlier inference on the possible values 
of $\alpha$ for the physical consistency. Because of these in our rest of the 
study, we consider the value of $\alpha > 0.6$. Figure \ref{02} 
shows the variation of the amplitude and damping of the QNMs 
with variation in the string parameter $a$ for three values of $l$. From the 
figure, it is clear that with increasing $a$, the amplitude of the QNM 
frequency decreases in such a way that the decrease is very minimal for lower 
$l$ and noticeable for higher $l$. The damping, on the other hand, shows 
a drastic increase with increasing $a$ values, but for higher values of 
$l$ such as $l=3$, the damping increases slightly and again decreases with 
increasing $a$ values. 
Figure \ref{03} shows the variation of the QNMs with charge $q$. The first 
plot of this figure shows the variation of the amplitude with $q$, showing 
that with increasing charge values, the amplitude also increases. It is also 
noteworthy that higher multipole $l$ also results in higher amplitude values. 
The damping is mostly higher for smaller $l$ values and the variation of 
damping with respect to $q$ increases with increasing charge values up to 
$q=0.7$. After this, there is a drastic decrease in damping as $q$ further 
increases showing the gradually decreasing difference of damping due to 
different $l$ values. Figure \ref{04} represents the variation of QNMs with 
Rastall parameter $\beta$, where we can see that the amplitude of the QNMs 
decreases with $\beta$  very slowly. Although this is not very clear 
from the figure, the tabular form of QNM frequencies (Table \ref{tab01}) 
clearly shows this trend. The damping decreases with increasing $\beta$ 
for the various values of $l$ as shown in the plot.

The QNMs for the black hole for various combinations of model parameters have 
been shown in Table \ref{tab01}, where we also show the calculated approximate 
errors associated with the 6th order WKB method. The process of error 
estimation has been adopted from \cite{50} where the authors have suggested 
the formula,
\begin{equation}
\Delta_{6}=\frac{1}{2}\Big|W\!K\!B_7 - W\!K\!B_5\Big|.
\label{b7}
\end{equation}
Here $\Delta_6$ represents the error associated with the 6th order WKB value 
of QNMs, $W\!K\!B_7$ and $W\!K\!B_5$ respectively means the 7th and 5th order 
QNM values. It is noteworthy that the WKB method produces reliable results 
for the $n<l$ regime only and its accuracy increases for higher $l$ as is
evident from our table also. For $n=0$, multipoles $l=1$ and $l=2$ are 
chosen and QNMs along with errors associated are shown in this Table 
\ref{tab01}. The first four sections in the table are for $l=1$. In the 
first section, we keep $a$, $\alpha$ and $q$ fixed and change $\beta$ 
to see its effect on the QNMs. It is clear that the amplitude or the real part 
of QNMs decreases with increasing $\beta$ and a similar trend is also seen in 
the imaginary part or the damping part. The error estimation is around 
$10^{-4}$. The second section shows the variation in $q$ while keeping $a$, 
$\alpha$ and $\beta$ fixed. Here, we can see that with an increase in $q$, 
the amplitude increases while the damping increases initially but later 
decreases for higher $q$, with error estimation about $10^{-4}$. 
In the third section, we fix $\beta$, $\alpha$ and $q$ and vary the string 
parameter $a$. We observe that the amplitude decreases while damping increases 
with an increase in $a$. The estimated error is about $10^{-5}$.
In the fourth section, we vary the $\alpha$ and fix the other parameters. 
It is seen that both the amplitude as well as damping decreases with 
increasing $\alpha$. Here, we have an error estimation of about $10^{-4}$. 
A Similar setup is presented in the lower sections but 
with $l=2$. Here the almost same trend is seen as in the case of $l=1$ but with 
higher QNM amplitudes, which is almost twice that for the $l=1$ case. In this 
case, the magnitude of damping is almost equal to that for the earlier 
case. The error estimation in this case decreases to $\sim 10^{-6}$. Thus it 
is quite clear that increasing the multipole number $l$ leads to more accuracy 
in WKB results. As a future scope, we can compare various other numerical and 
analytical methods of calculating QNMs for our particular black hole in 
modified Rastall gravity, surrounded by clouds of strings. 
\begin{table}[h!]
\caption{6th order WKB QNMs of the black hole specified by the 
metric function \eqref{a19} for the multipoles $l=1,2$ with $n=0$ and for 
different values of the model parameters. The estimated errors associated with 
the WKB results have been shown.}
\vspace{5mm}
\centering
\begin{tabular}{c@{\hskip 5pt}c@{\hskip 10pt}c@{\hskip 10pt}c@{\hskip 10pt}c@{\hskip 10pt}c@{\hskip 10pt}c@{\hskip 10pt}c@{\hskip 10pt}c}
\hline \hline
\vspace{2mm}
& Multipole & $a$ & $\alpha$ & $\beta$ & $q$ & 6th order WKB QNMs & $\Delta_{6}$  \\
\hline
&\multirow{4}{4em}{$l=1$} & $0.01$ & $0.80$ & $0.01$ & $0.80$ & 0.319038 - 0.101938i  & $3.5059 \times 10^{-4}$ &\\ 
& & $0.01$ & $0.80$ & $0.05$ & $0.80$ & 0.309857 - 0.098262i  & $7.2185 \times 10^{-4}$ &\\
& & $0.01$ & $0.80$ & $0.10$ & $0.80$ & 0.290292 - 0.093340i & $5.6480 \times 10^{-5}$ &\\
& & $0.01$ & $0.80$ & $0.12$ & $0.80$ & 0.277566 - 0.090237i & $1.3723 \times 10^{-3}$ &\\
\hline
&\multirow{4}{4em}{$l=1$} & $0.001$ & 0.80 & $0.01$ & $0.50$ & 0.304581 - 0.099204i & $7.7780 \times 10^{-5}$ & \\
& & $0.001$ & 0.80 & $0.01$ & $0.70$ & 0.321046 - 0.099692i &  $4.1684 \times 10^{-5}$ &\\
& & $0.001$ & 0.80 & $0.01$ & $0.80$ & 0.333653 - 0.099347i & $1.0797 \times 10^{-4}$ &\\
& & $0.001$ & 0.80 & $0.01$ & $0.90$ & 0.351315 - 0.097390i & $1.3079\times 10^{-3}$ & \\
\hline
&\multirow{4}{4em}{$l=1$} & $0.001$ & 0.80 & $0.01$ & $0.50$ & 0.304581 - 0.099204i & $7.7780 \times 10^{-5}$ &\\ 
&& $0.005$ & 0.80 & $0.01$ & $0.50$ & 0.296683 - 0.100816i & $3.3699 \times 10^{-5}$ &\\
&& $0.010$ & 0.80 & $0.01$ & $0.50$ & 0.285134 - 0.102889i & $2.5937 \times 10^{-4}$ &\\
&& $0.020$ & 0.80 & $0.01$ & $0.50$ & 0.258318 - 0.103706i & $1.2794 \times 10^{-4}$ & \\
\hline
&\multirow{4}{4em}{$l=1$} & $0.01$ & 0.70 & $0.01$ & $0.80$ & 0.322953 - 0.105406i  & $1.9497 \times 10^{-4}$ &\\ 
&& $0.01$ & 0.75 & $0.01$ & $0.80$ & 0.320730 - 0.103206i  & $7.4452 \times 10^{-4}$ &\\
&& $0.01$ & 0.80 & $0.01$ & $0.80$ & 0.319038 - 0.101938i & $3.5060 \times 10^{-4}$ &\\
&& $0.01$ & 0.90 & $0.01$ & $0.80$ & 0.316019 - 0.100356i & $5.4711 \times 10^{-4}$ &\\
\hline
&\multirow{4}{4em}{$l=2$} & $0.01$ & 0.80 & $0.01$ & $0.80$ & 0.525770 - 0.099130i  & $5.3032 \times 10^{-6}$ &\\ 
& & $0.01$ & 0.80 & $0.05$ & $0.80$ & 0.514233 - 0.095527i  & $9.0525 \times 10^{-6}$ &\\
& & $0.01$ & 0.80 & $0.10$ & $0.80$ & 0.485430 - 0.090238i & $9.2432 \times 10^{-6}$ &\\
& & $0.01$ & 0.80 & $0.12$ & $0.80$ & 0.466354 - 0.086931i & $1.9280 \times 10^{-6}$ &\\
\hline
&\multirow{4}{4em}{$l=2$} & $0.01$ & 0.80 & $0.01$ & $0.50$ & 0.470181 - 0.098794i & $1.5440 \times 10^{-5}$ & \\
& & $0.01$ & 0.80 & $0.01$ & $0.70$ & 0.501907 - 0.099424i & $1.0620 \times 10^{-5}$ &\\
& & $0.01$ & 0.80 & $0.01$ & $0.80$ & 0.525770 - 0.099130i & $5.3032 \times 10^{-6}$ &\\
& & $0.01$ & 0.80 & $0.01$ & $0.90$ & 0.558772 - 0.097247i & $3.3164 \times 10^{-6}$ & \\
\hline
&\multirow{4}{4em}{$l=2$} & $0.001$ & 0.80 & $0.01$ & $0.50$ & 0.502511 - 0.098143i & $9.3315 \times 10^{-6}$ &\\ 
&& $0.005$ & 0.80 & $0.01$ & $0.50$ & 0.488420 - 0.098703i  & $1.0838 \times 10^{-6}$ &\\
&& $0.010$ & 0.80 & $0.01$ & $0.50$ & 0.470181 - 0.098794i & $4.4192 \times 10^{-6}$ &\\
&& $0.020$ & 0.80 & $0.01$ & $0.50$ & 0.432185 - 0.096708i & $1.0715 \times 10^{-5}$ & \\
\hline
&\multirow{4}{4em}{$l=2$} & $0.01$ & 0.70 & $0.01$ & $0.80$ & 0.527277 - 0.101960i & $5.4031 \times 10^{-5}$ &\\ 
&& $0.01$ & 0.75 & $0.01$ & $0.80$ & 0.526872 - 0.100228i & $2.5585 \times 10^{-5}$ &\\
&& $0.01$ & 0.80 & $0.01$ & $0.80$ & 0.525770 - 0.099130i & $5.3032 \times 10^{-6}$ &\\
&& $0.01$ & 0.90 & $0.01$ & $0.80$ & 0.522562 - 0.097652i & $2.9765 \times 10^{-6}$ & \\
\hline \hline \vspace{4mm}
\end{tabular}
\label{tab01}
\end{table}

\begin{figure}[h!]
\includegraphics[scale=0.3]{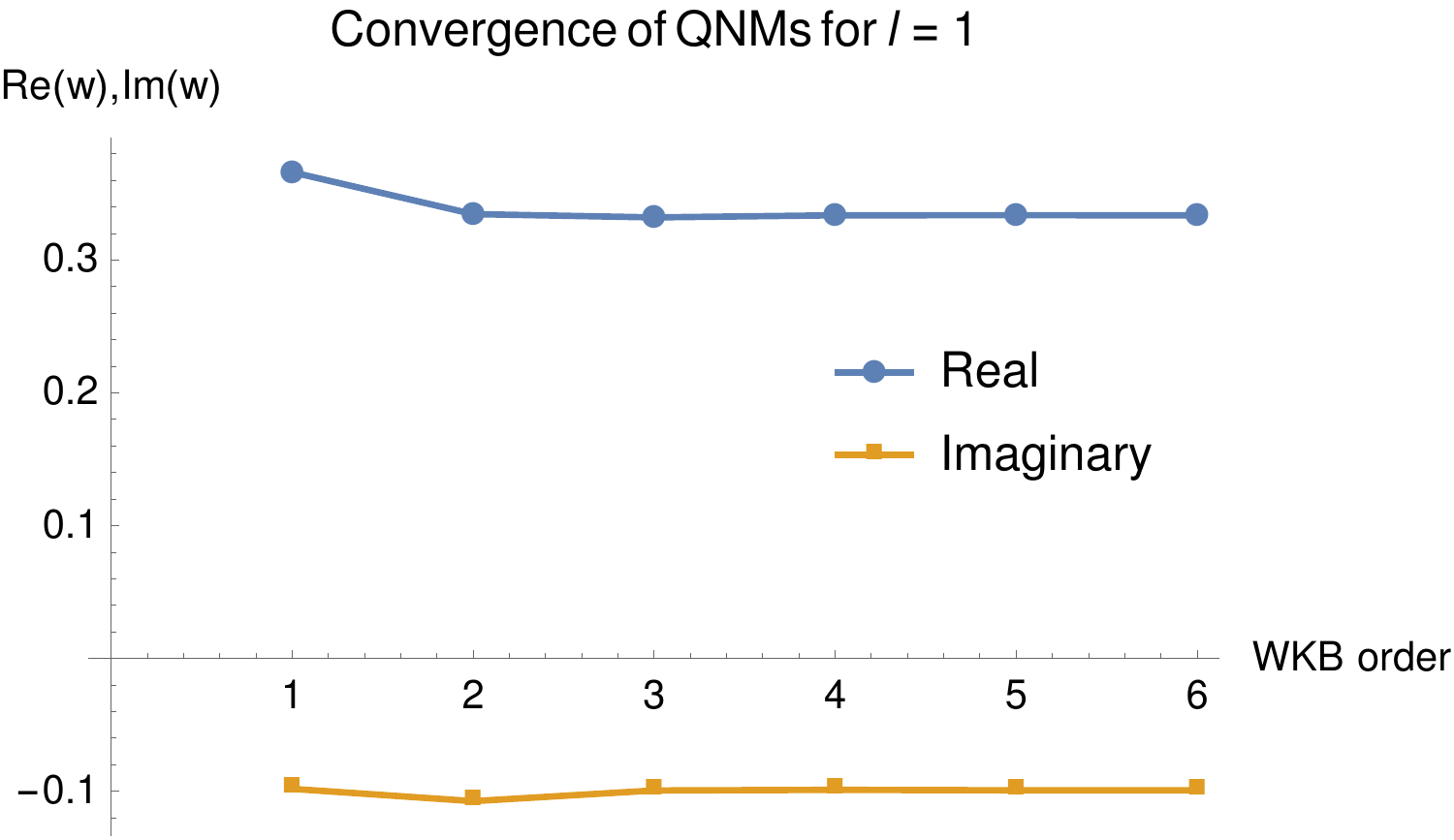}\hspace{0.5cm}
\includegraphics[scale=0.3]{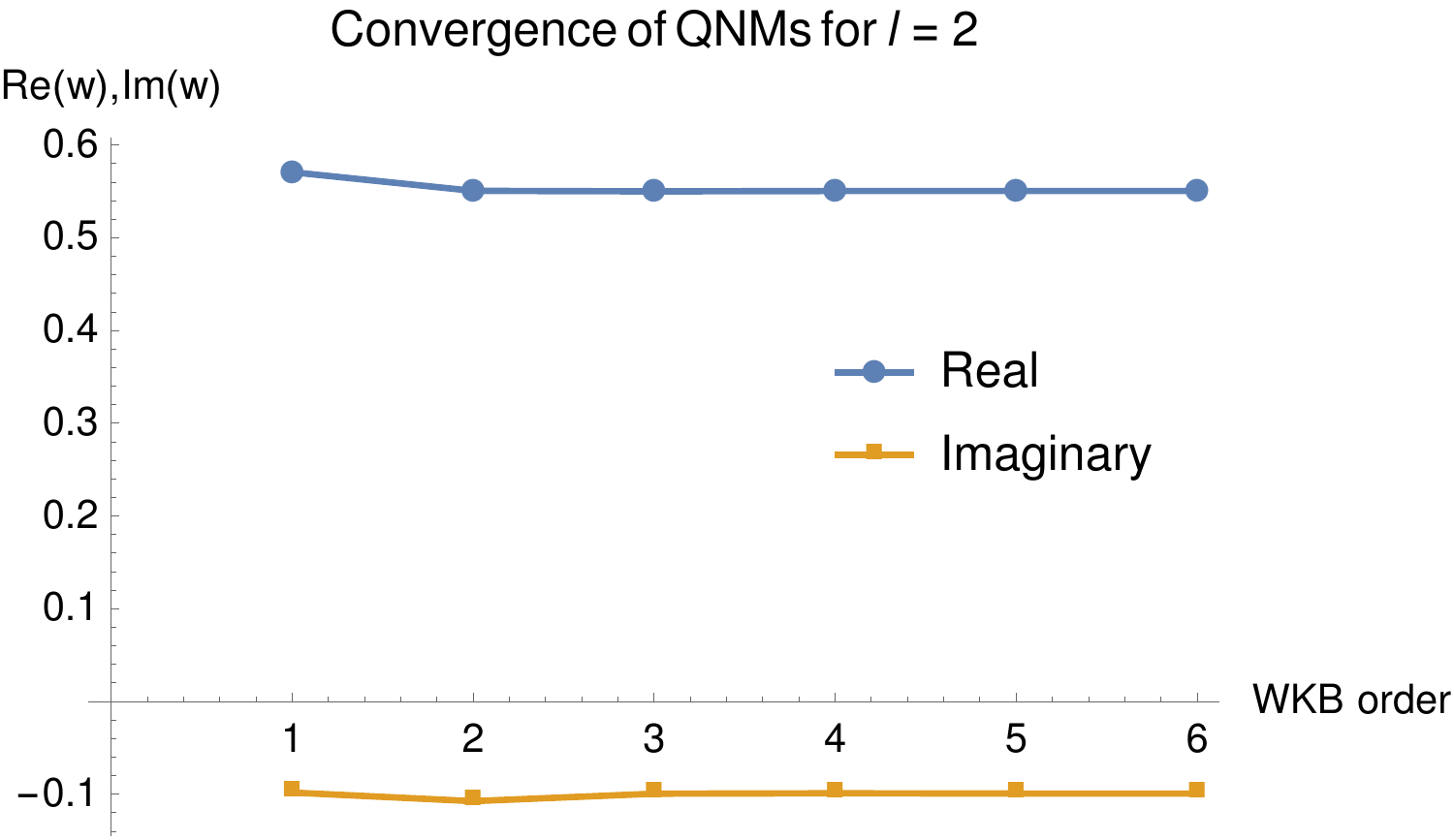}
\caption{Convergence of quasinormal modes of oscillations of the black hole 
described by the metric function \eqref{a19} for different WKB orders. For this 
figure we used $l=1$, $a=0.001$, $q=1$, $\alpha=0.80$ and $\beta=0.01$.}
\label{05}
\end{figure}

We also plot the convergence of the QNMs of various WKB orders in Figure 
\ref{05}. We have shown the convergence of QNMs up to 6th WKB order calculation.%

\begin{figure}[h!]
\centerline{
\includegraphics[scale=0.35]{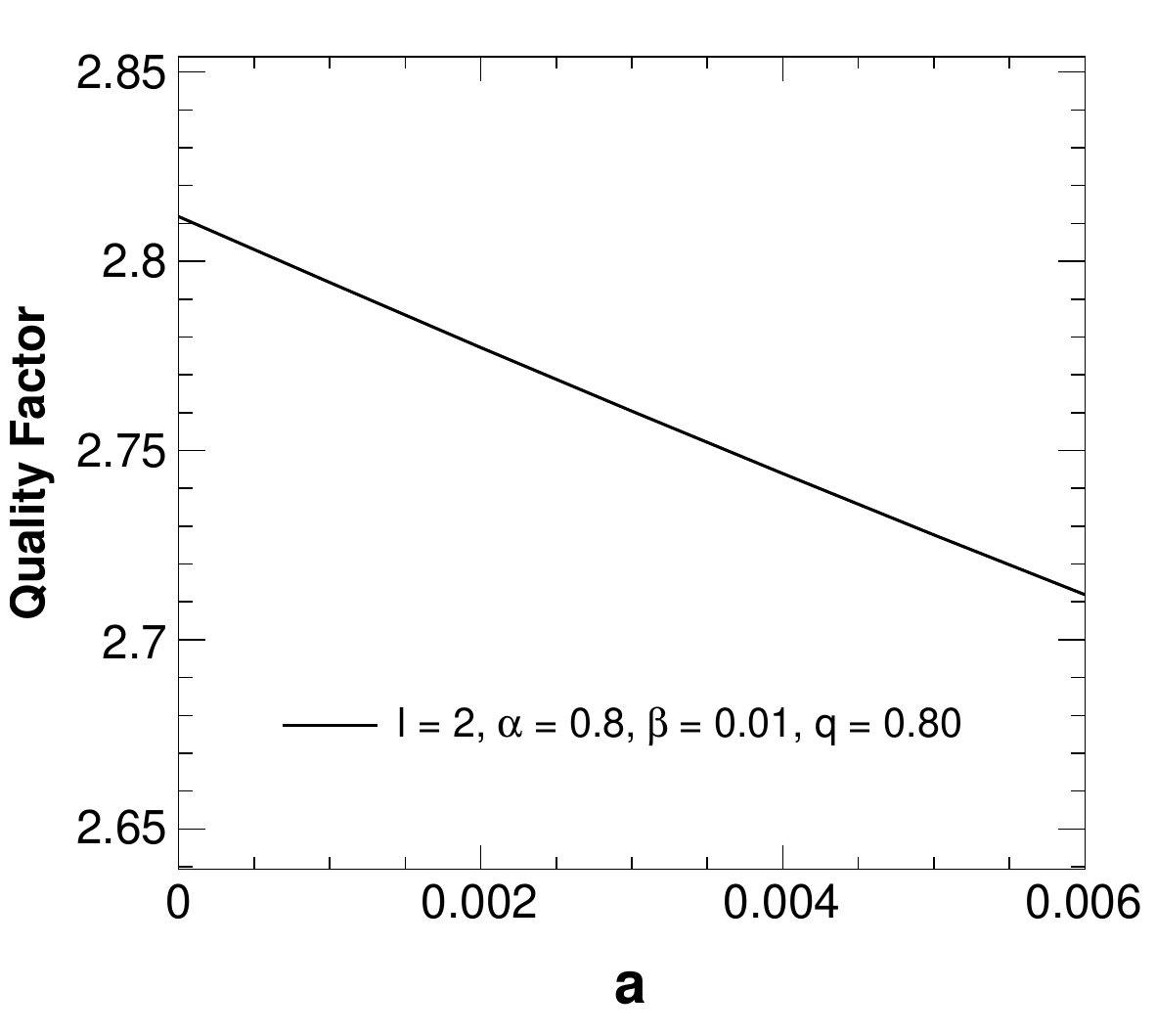}\hspace{0.5cm}
\includegraphics[scale=0.35]{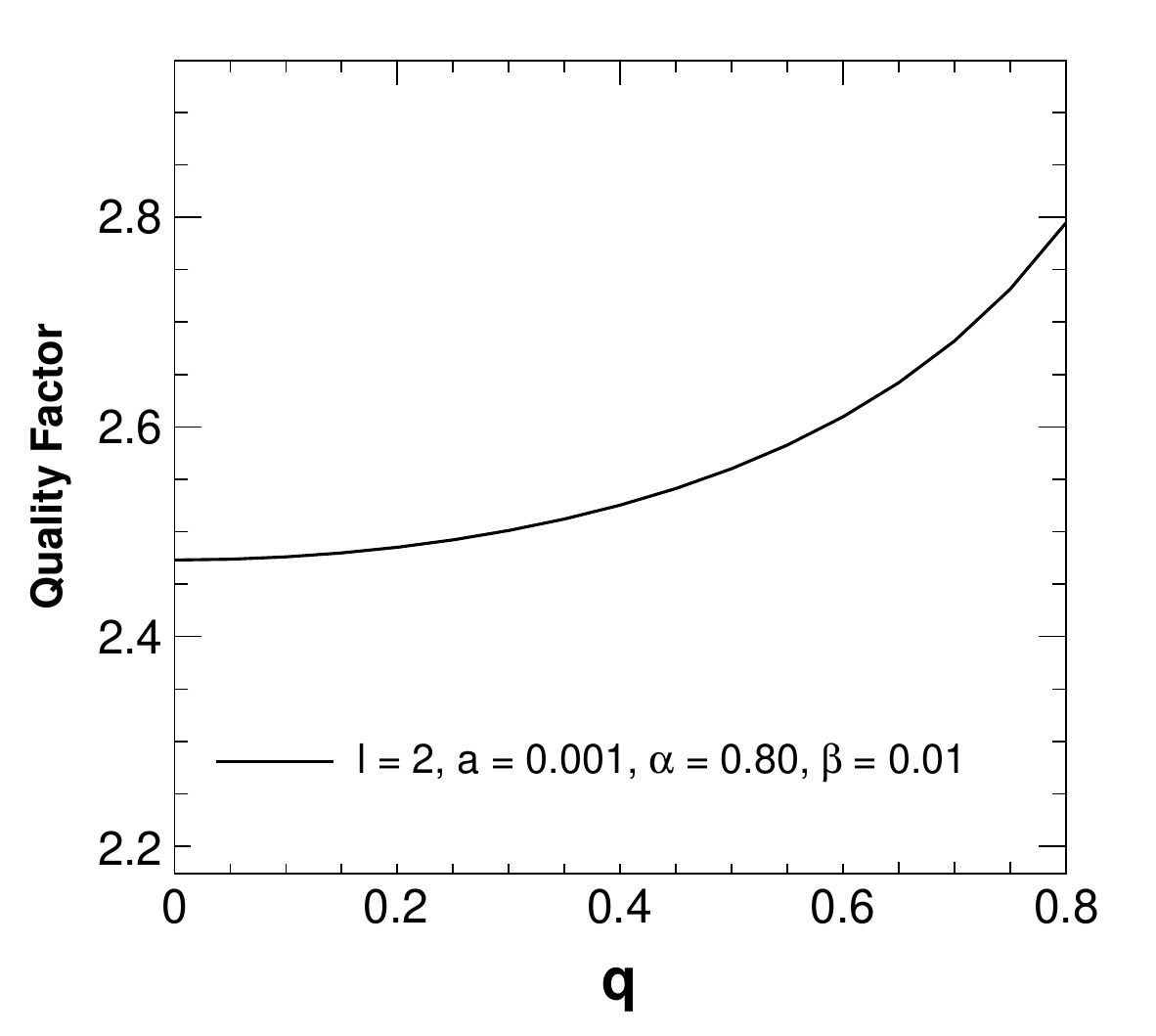}}\vspace{0.3cm}
\centerline{
\includegraphics[scale=0.35]{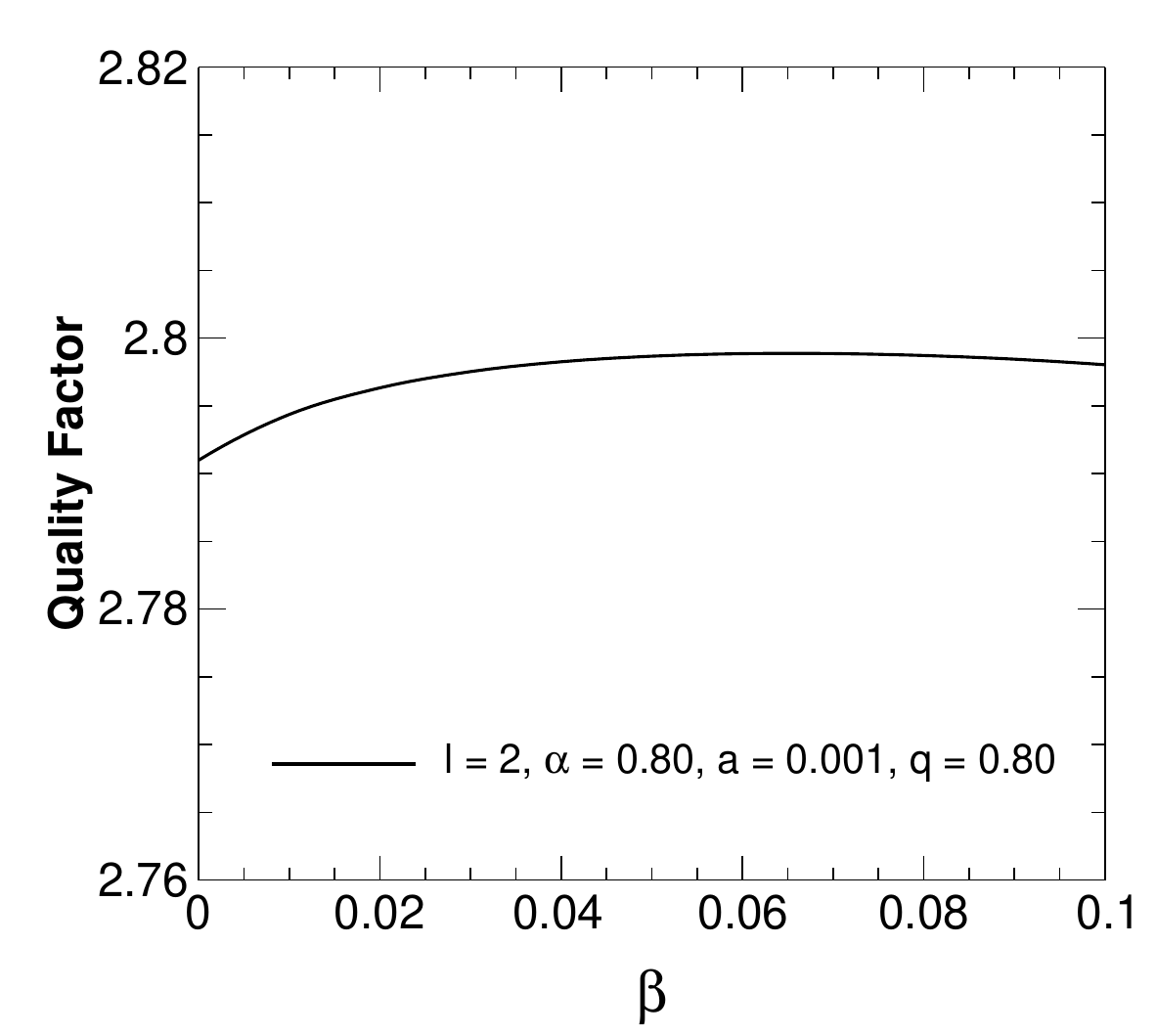}\hspace{0.5cm}
\includegraphics[scale=0.35]{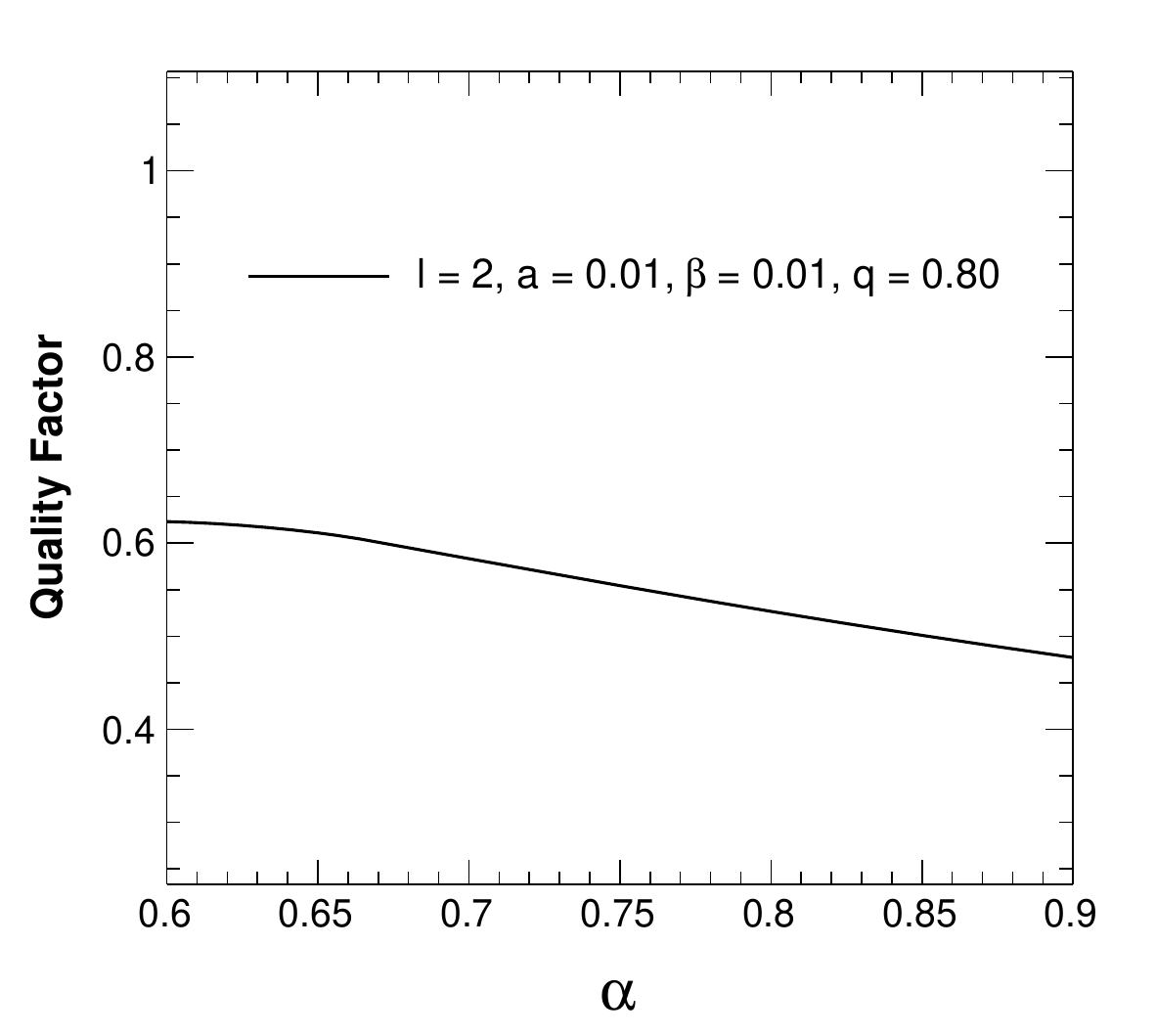}}
\vspace{-0.2cm}
\caption{Variation of quality factor as a function of $a$, $q$, $\beta$ and 
$\alpha$. The values of the parameters used have been mentioned in the plot 
itself.}
\label{061}
\end{figure}

The quality factor is a good means of showcasing the strength of a wave of 
oscillations versus its damping. It is basically a dimensionless quantity and 
demonstrates how under-damped the wave is. The more the quality factor, the 
more will be the strength of oscillation compared to damping. Mathematically 
we express the quality factor by the formula:
\begin{equation}
\text{Quality Factor} = \frac{Re(\omega)}{2* Im(\omega)}.
\label{b8}
\end{equation}
We plot the quality factor of QNMs for our black hole solution with respect 
to charge $q$, string parameter $a$ and Rastall parameter $\beta$ and 
parameter $\alpha$ in Figure \ref{061}. It is clear from the figure that the 
quality factor decreases with an increase in $a$ and $\alpha$ while the quality 
factor increases gradually with an increase in $q$. Concerning $\beta$, this
factor first increases slowly, then decreases minutely with an increase in 
$\beta$. This means that the black hole system becomes overdamped 
with increasing values of $a$ and $\alpha$.

\section{Temperature of the black hole and its characteristics}\label{sec4}
The temperature of a black hole is an important characteristic, which we want 
to analyse for our model. From the metric function \eqref{a19}, we can get 
the event horizon radius for our black hole by the condition 
$f(r\rightarrow r_H)=0$, where $r_H$ implies the horizon radius of the black
hole. In the general Schwarzschild case, we encounter a comparatively simple 
expression but in our case, we cannot get an analytical expression directly 
for the horizon radius because of the complicated nature of the metric function 
and hence we shall follow some other methods accordingly. To this end, we 
first derive the surface gravity of the black hole using the relation 
\cite{26},
\begin{equation}
\kappa_{g}=\frac{1}{2}\frac{df(r)}{dr}\Big|_{r\,=\,r_H}.
\label{c1}
\end{equation}
From the numerical calculations we encounter three horizons in the case of our 
metric for a specific set of values of model parameters (see Figure \ref{01}). 
The innermost horizon is called the inner or Cauchy horizon, the next horizon 
is referred to as the outer or event horizon and the last one is the outermost 
or the string horizon respectively. Using the expression of the surface 
gravity, we can derive the Hawkings temperature by using the relation \cite{26},
\begin{equation}
T_{BH}=\frac{\kappa_{g}}{2\pi}=\frac{1}{4\pi}\frac{df(r)}{dr}\Big|_{r\,=\,r_H}.
\label{c2}
\end{equation}
Here horizon radius $r_{H}\ $ has three values, viz., $r_{-}$, 
$r_{+}$ and $r_{c}$ which are respectively the inner horizon or Cauchy 
horizon, the event horizon and the string horizon. We follow Ref.\ \cite{28-1} 
to plot Hawking temperature with respect to $r_{H}$ which encompasses all three 
horizons. Before that, we numerically compute the Hawking temperature and its 
variation with changing values of model parameters along with the horizon 
radius values, which are shown in Table \ref{tab02}. In this table, we show 
the variation of horizon radii and temperature for various values of model 
parameters. It is clear from this table that $a$ and $\alpha$ parameters 
influence the second and third horizon, while not impacting the first horizon. 
Parameter $q$ impacts both the first and second horizon while not impacting 
third horizon. Finally, $\beta$ affects the third horizon, meaning that with 
increasing $\beta$, the third horizon radius increases and temperature 
decreases.

\begin{table}[h!]
\caption{Hawking temperature for the three horizon radii calculated numerically for various values of model parameters.}
\vspace{5mm}
\centering
\begin{tabular}{c@{\hskip 10pt}c@{\hskip 10pt}c@{\hskip 10pt}c@{\hskip 10pt}c@{\hskip 10pt}c@{\hskip 10pt}c@{\hskip 10pt}c@{\hskip 10pt}c@{\hskip 10pt}c}
\hline \hline
\vspace{2mm}
  $a$ & $\alpha$ & $\beta$ & $q$ & $r_{-}$ & $r_{+}$ & $r_c$ & $T_{r_{-}}$ & $T_{r_{+}}$ & $T_{r_c}$ \\
\hline
 0.001 & 0.6 & 0.01 & 0.80 & 0.4000 & 1.0200 & 6.3136 & -0.59683 & 0.0372 & -0.0447\\
 
 0.010 & 0.6 & 0.01 & 0.80 & 0.4000 & 1.7316 & 2.2235 & -0.59687 & 0.0187 & -0.0222 \\

 0.010 & 0.7 & 0.01 & 0.80 & 0.4000 & 1.6105 & 5.5805 & -0.59686 & 0.0360 & -0.0328 \\
 
0.001 & 0.6 & 0.01 & 0.90 & 0.5641 & 1.4364 & 6.3211 & -0.21801 & 0.0336 & -0.0451\\

0.001 & 0.6 & 0.10 & 0.90 & 0.5641 & 1.4382 & 19.3570 & -0.21810 & 0.0335 & -0.0093\\

\hline \hline \vspace{4mm}

\end{tabular}
\label{tab02}

\end{table}

\begin{figure}
\includegraphics[scale=0.365]{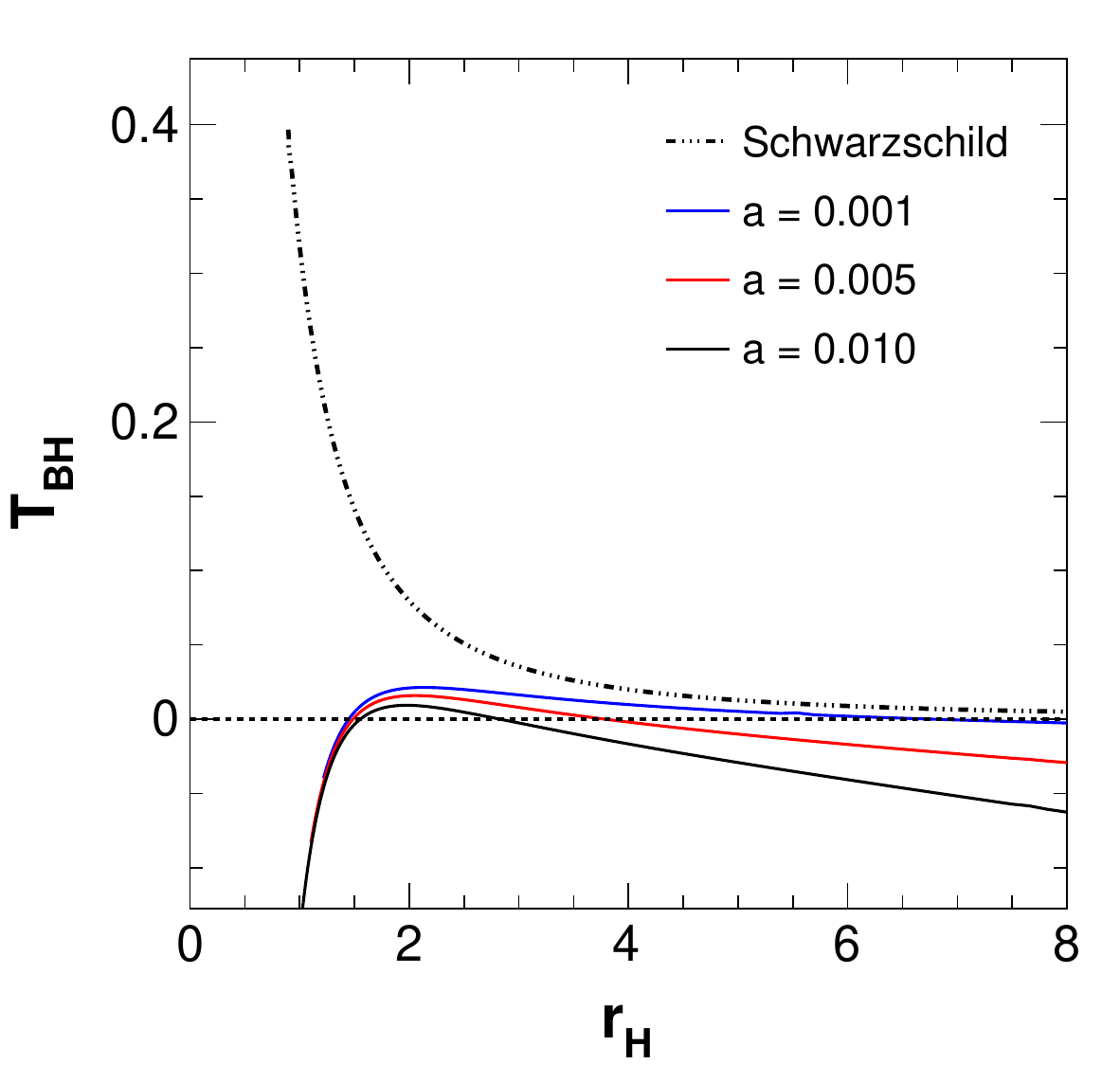}\hspace{0.5cm}
\includegraphics[scale=0.365]{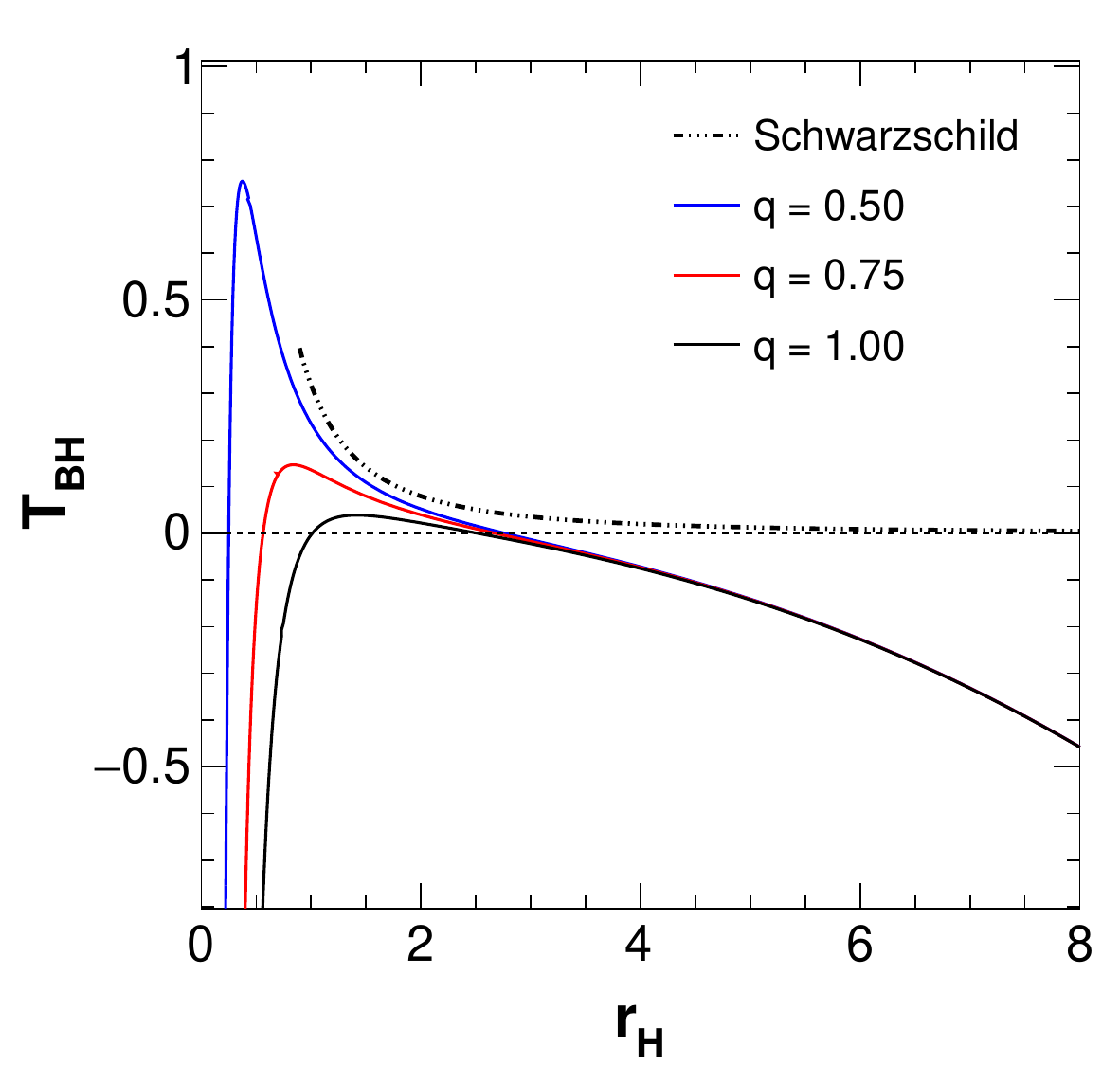}\vspace{0.3cm}
\includegraphics[scale=0.365]{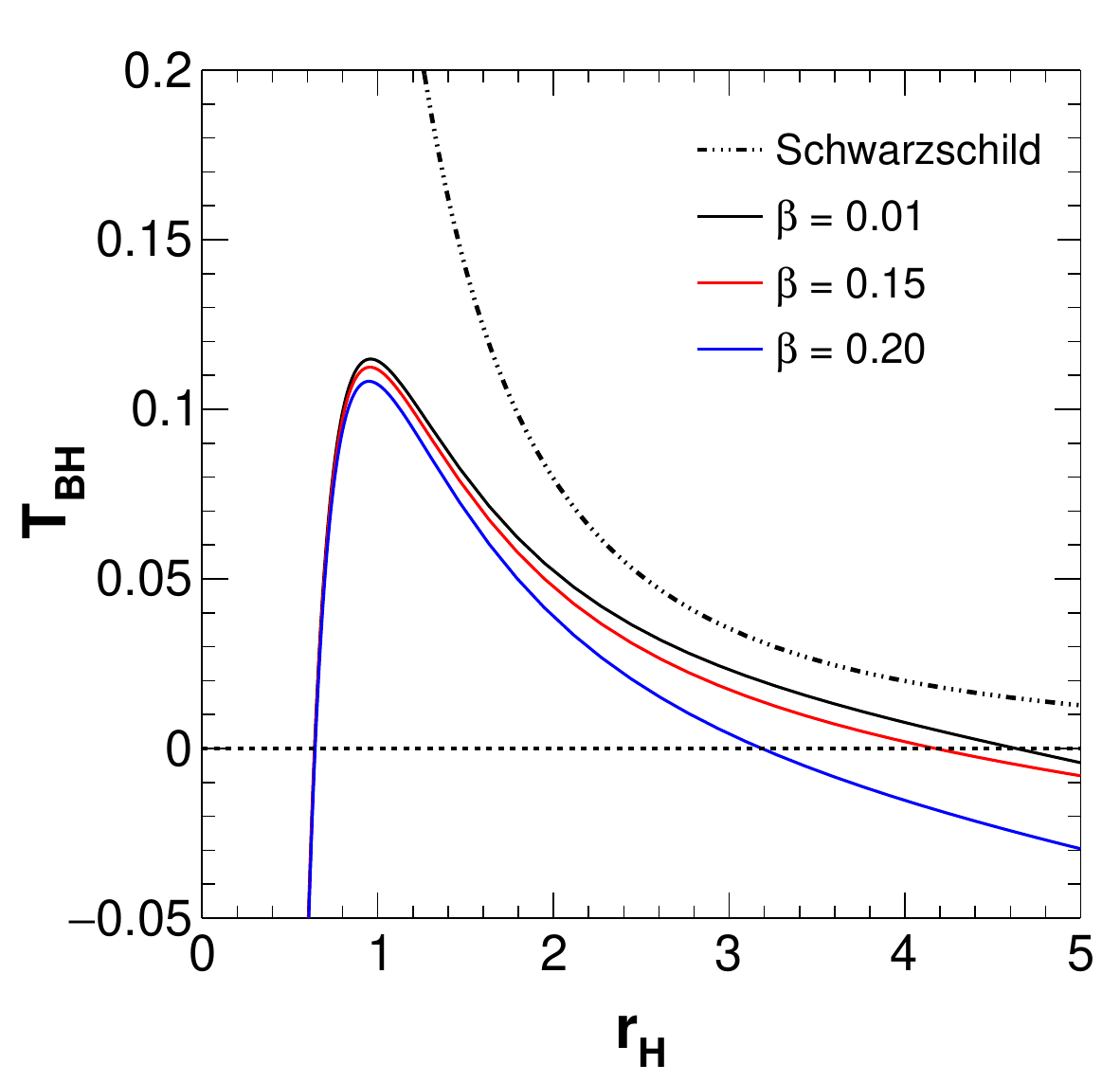}\hspace{0.5cm}
\includegraphics[scale=0.365]{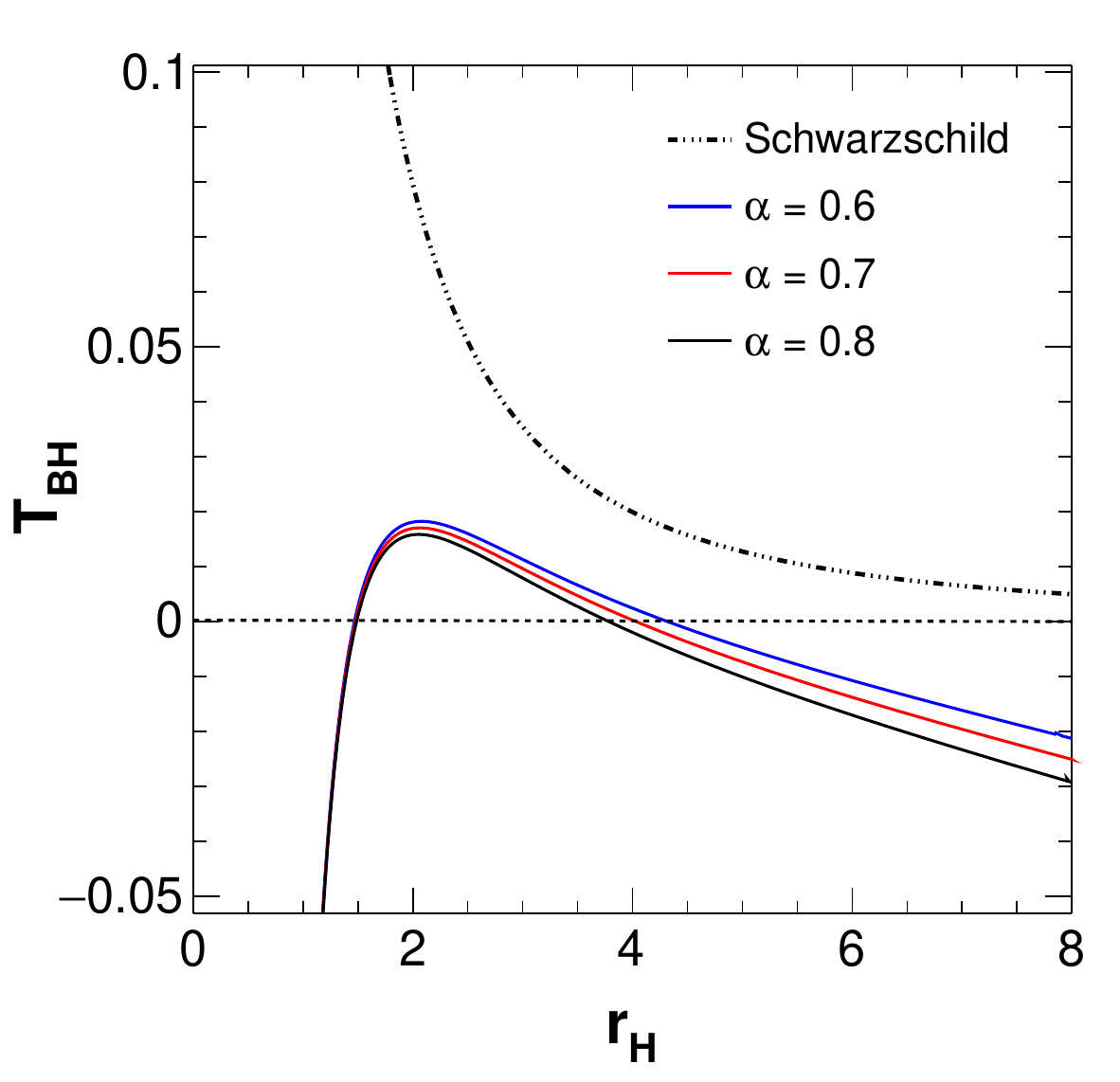}\vspace{0.3cm}
\vspace{-0.2cm}
\caption{Temperature versus horizon radius of the black hole for different 
values of the model parameters. The first plot is for $q=1.20$, $\beta=0.15$, 
$\alpha=0.80$, the second plot is for $a=0.05$, $\beta=0.01$, $\alpha=0.80$, 
the third plot uses $a=0.005$, $q=0.80$, $\alpha=0.80$ and the fourth one 
uses $a=0.005$, $\beta=0.01$, $q=1.20$.}
\label{06}
\end{figure}

Figure \ref{06} shows the variation of black hole temperature with horizon 
radius for different values of the parameters. The first plot represents 
temperature versus horizon radius for different values of the string parameter 
$a$. It is observed that with an increase in $a$, the graph deviates more from 
the ideal Schwarzschild case towards the negative temperature side and 
for $r_H < 1.5$, the temperature drops down to negative values. Similar is 
the case with higher horizon radius values where the temperature gradually 
goes towards the negative side. The second plot shows the variation of 
temperature with horizon radius for various values of charge $q$ and the 
third plot is shown for various values of the parameter $\beta$. 
Similarly, the fourth plot shows temperature variation curves for different 
$\alpha$ values. It 
is clear that for all the scenarios, at a very small horizon radius, the 
black hole becomes ultracold with negative temperatures which doesn't sound 
very physical but this has been encountered in research work before 
\cite{26,t1}. Similarly, the temperature of black holes with increasingly 
higher horizon radii becomes increasingly more negative.   

\section{Greybody factors and absorption cross section}\label{sec5}
\begin{figure}[h!]
\includegraphics[scale=0.35]{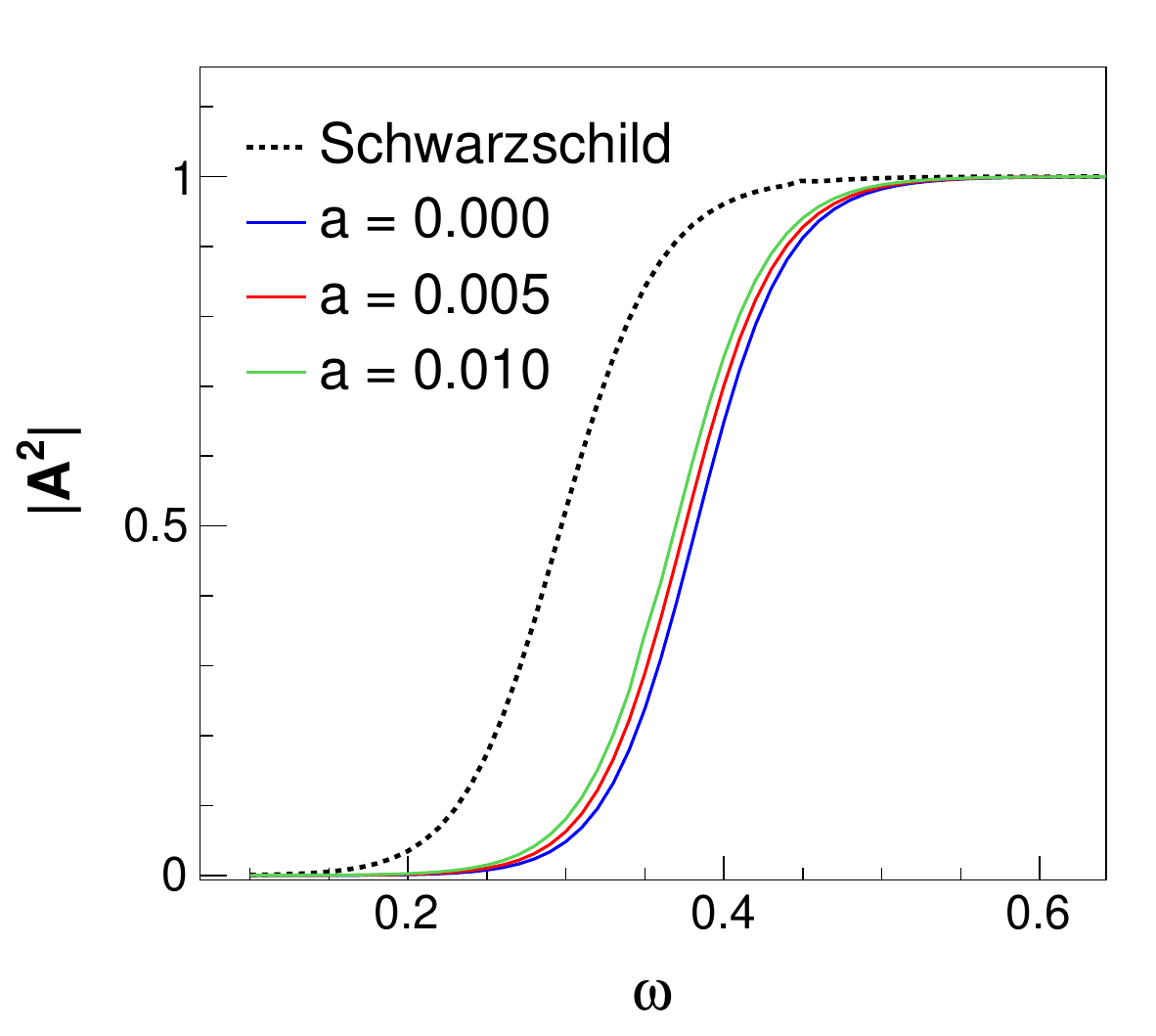}\hspace{0.5cm}
\includegraphics[scale=0.35]{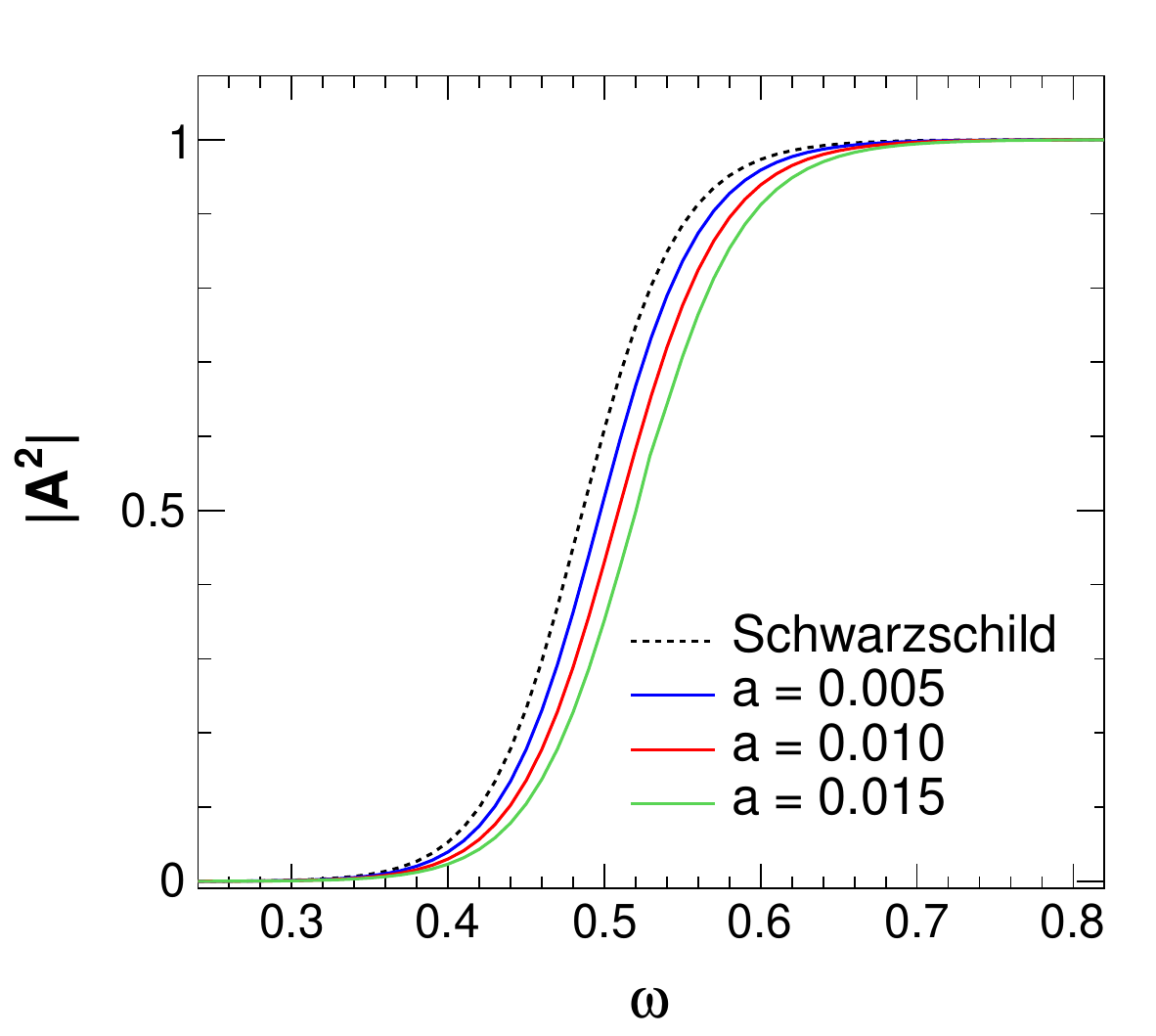}
\vspace{-0.2cm}
\caption{Greybody factor versus quasinormal mode frequency $\omega$ for 
different values of the model parameter $a$. The left plot is for multipole 
$l=1$ and the right plot is for multipole $l=2$. Here $q=1$, $\alpha=0.80$ and $\beta=0.01$ are used.}
\label{07}
\end{figure}

\begin{figure}[h!]
\includegraphics[scale=0.35]{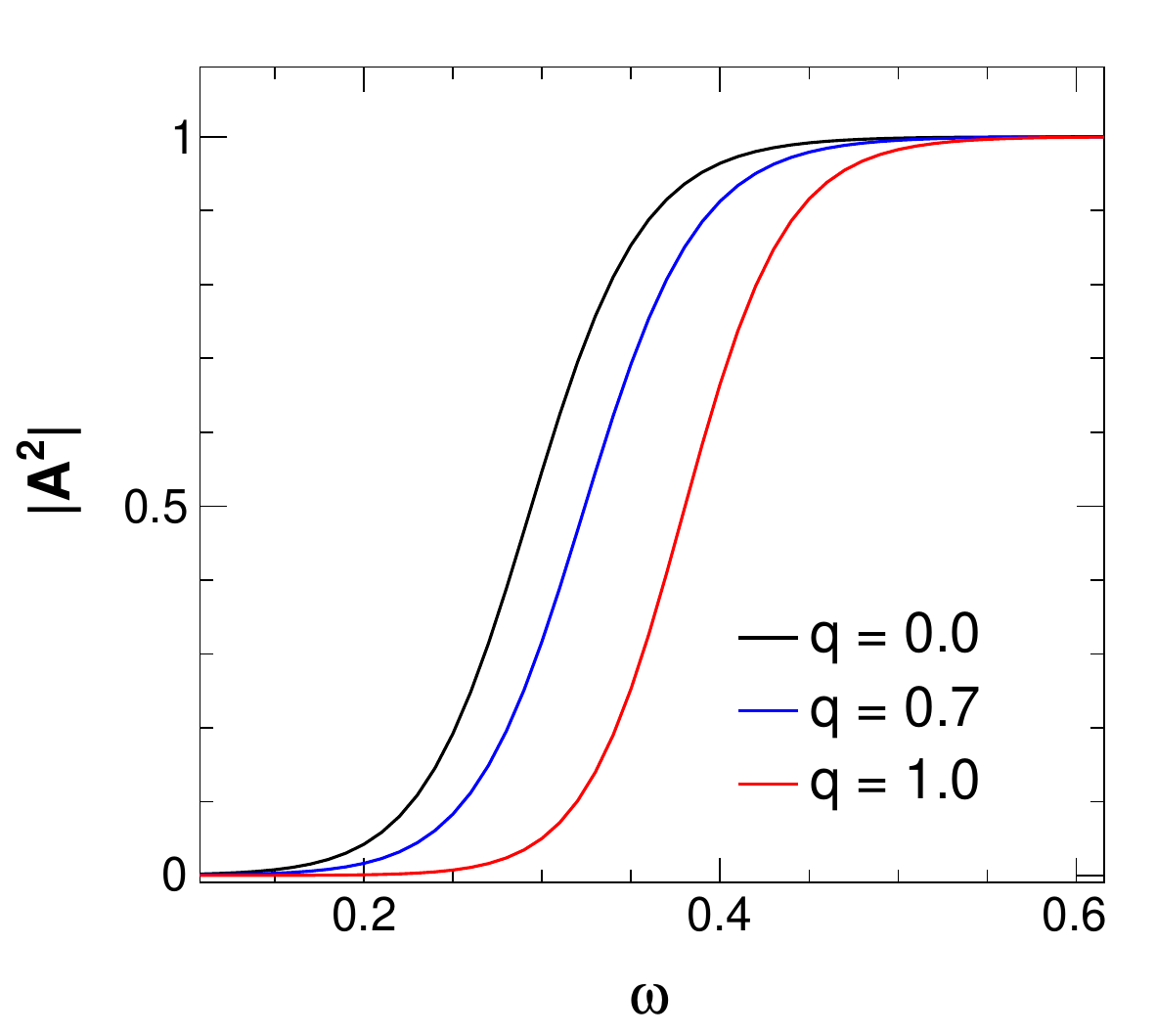}\hspace{0.5cm}
\includegraphics[scale=0.35]{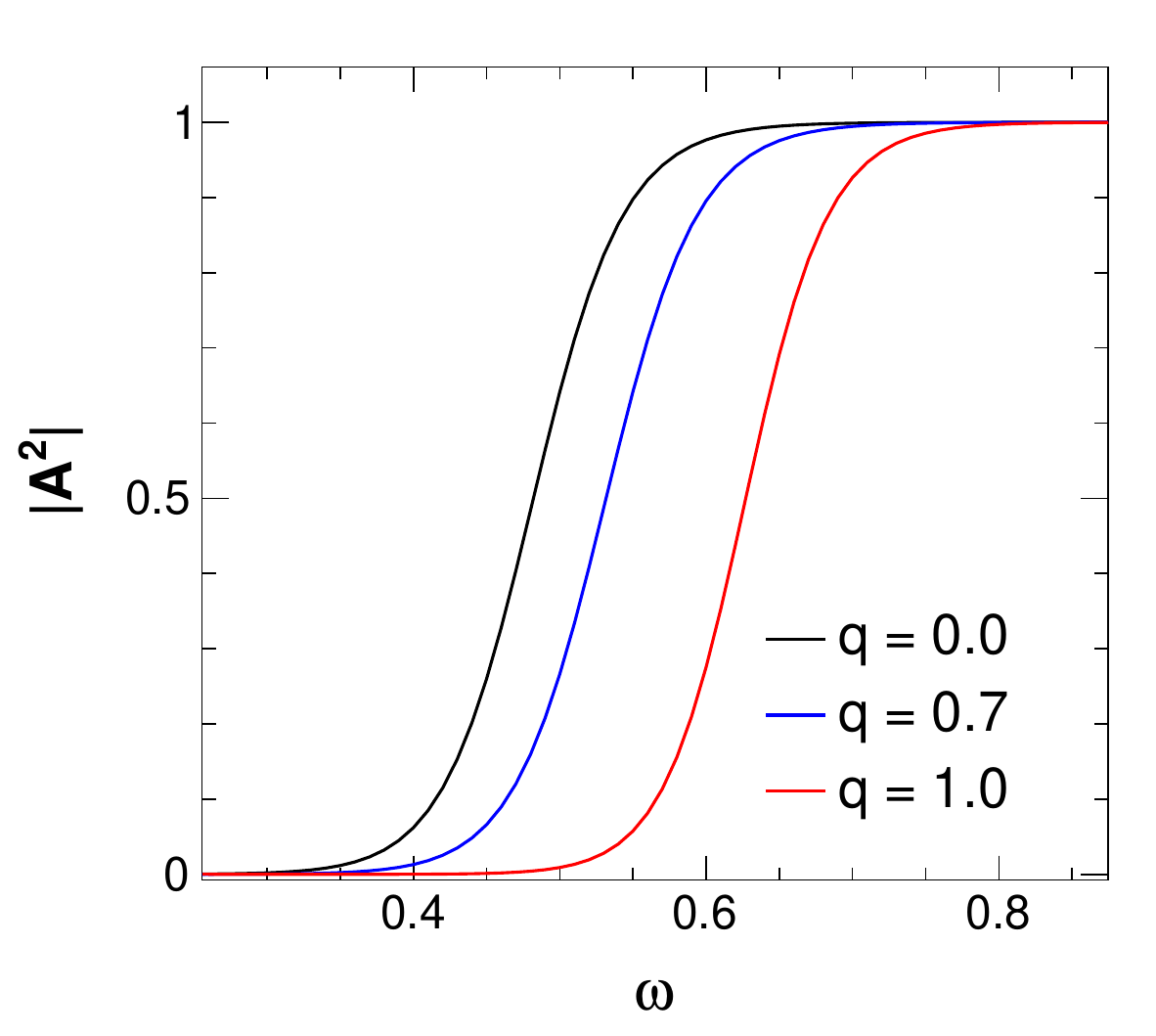}
\vspace{-0.2cm}
\caption{Greybody factor versus quasinormal mode frequency $\omega$ for 
different values of the model parameter $q$. The left plot is for multipole 
$l=1$ and the right plot is for multipole $l=2$. Here $a=0.001$, $\alpha=0.80$ 
and $\beta=0.01$ are used.}
\label{08}
\end{figure}

\begin{figure}[h!]
\includegraphics[scale=0.35]{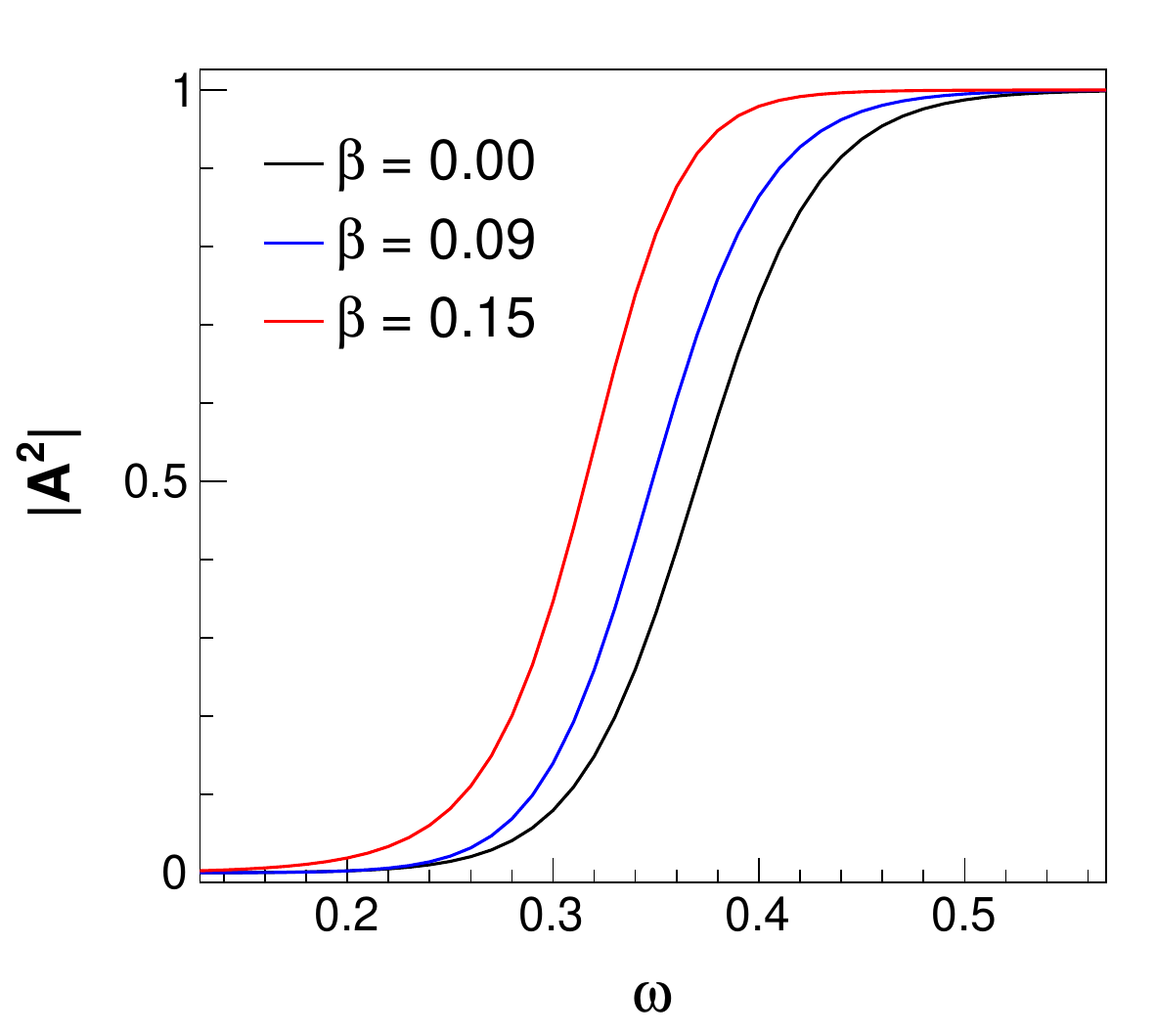}\hspace{0.5cm}
\includegraphics[scale=0.35]{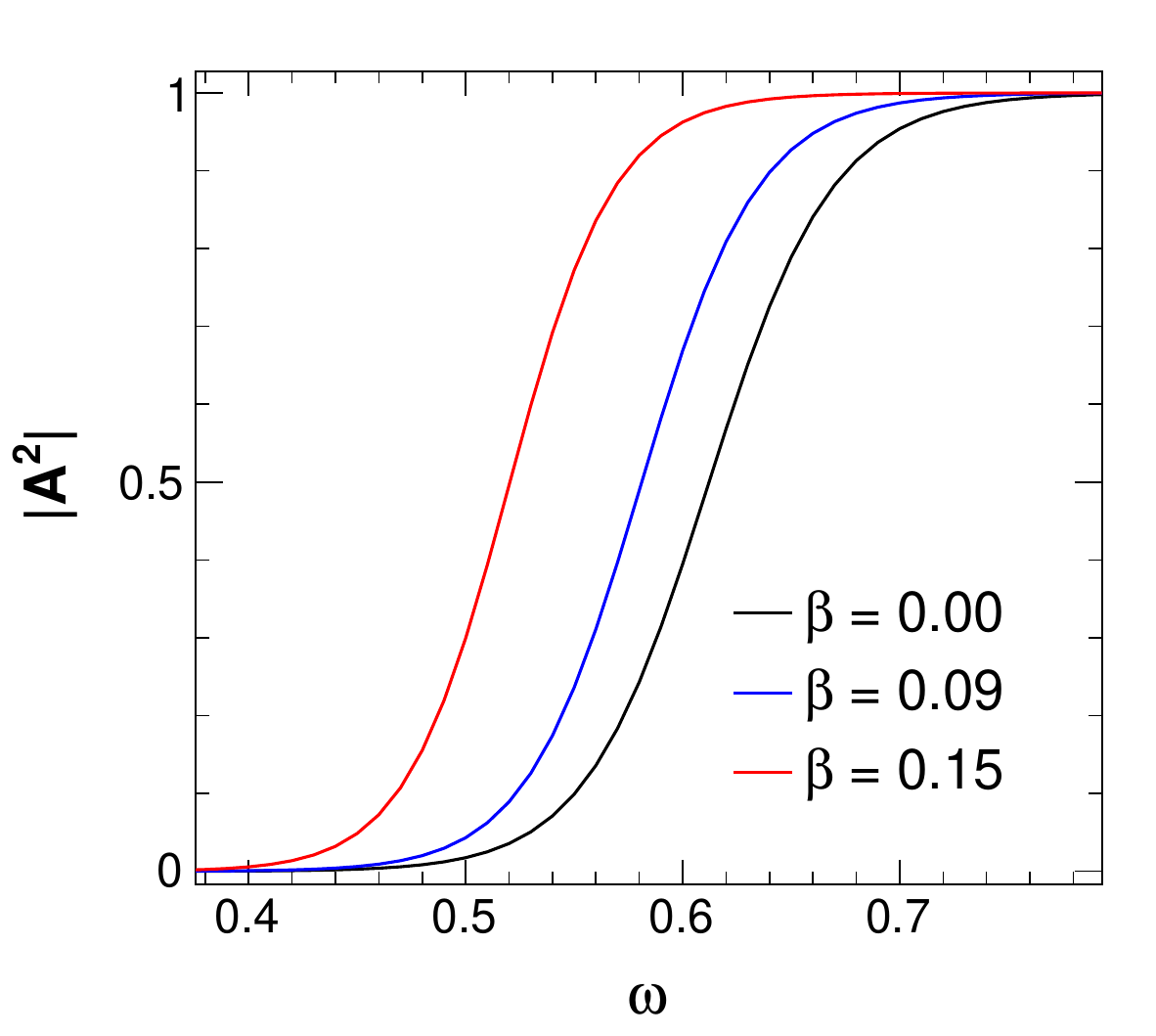}
\vspace{-0.2cm}
\caption{Greybody factor versus quasinormal mode frequency $\omega$ for 
different values of the model parameter $\beta$. The left plot is for 
multipole $l=1$ and the right plot is for multipole $l=2$. Here $q=1$, 
$\alpha=0.80$ and $a=0.01$ used.}
\label{09}
\end{figure}

\begin{figure}[h!]
\includegraphics[scale=0.34]{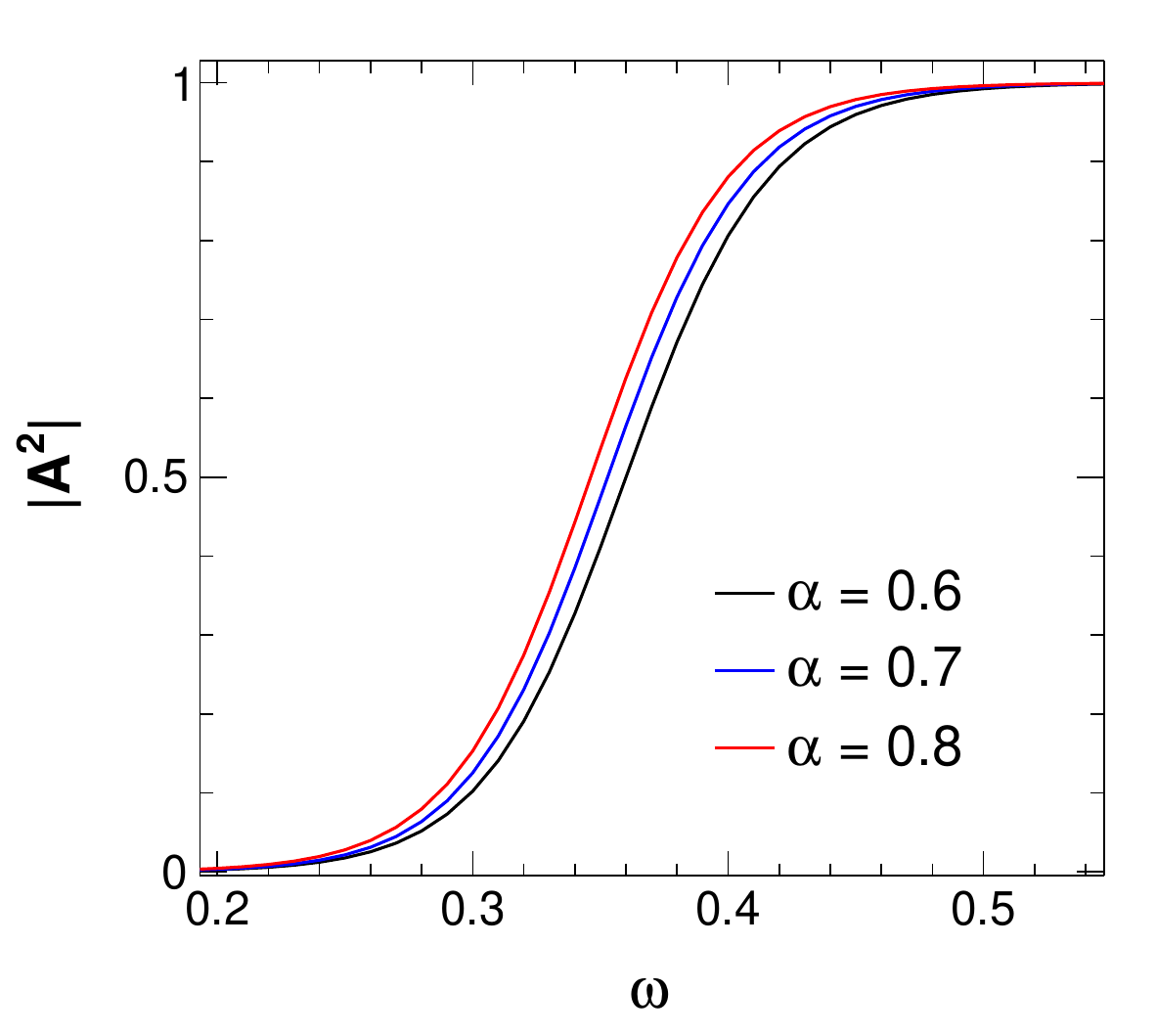}\hspace{0.5cm}
\includegraphics[scale=0.34]{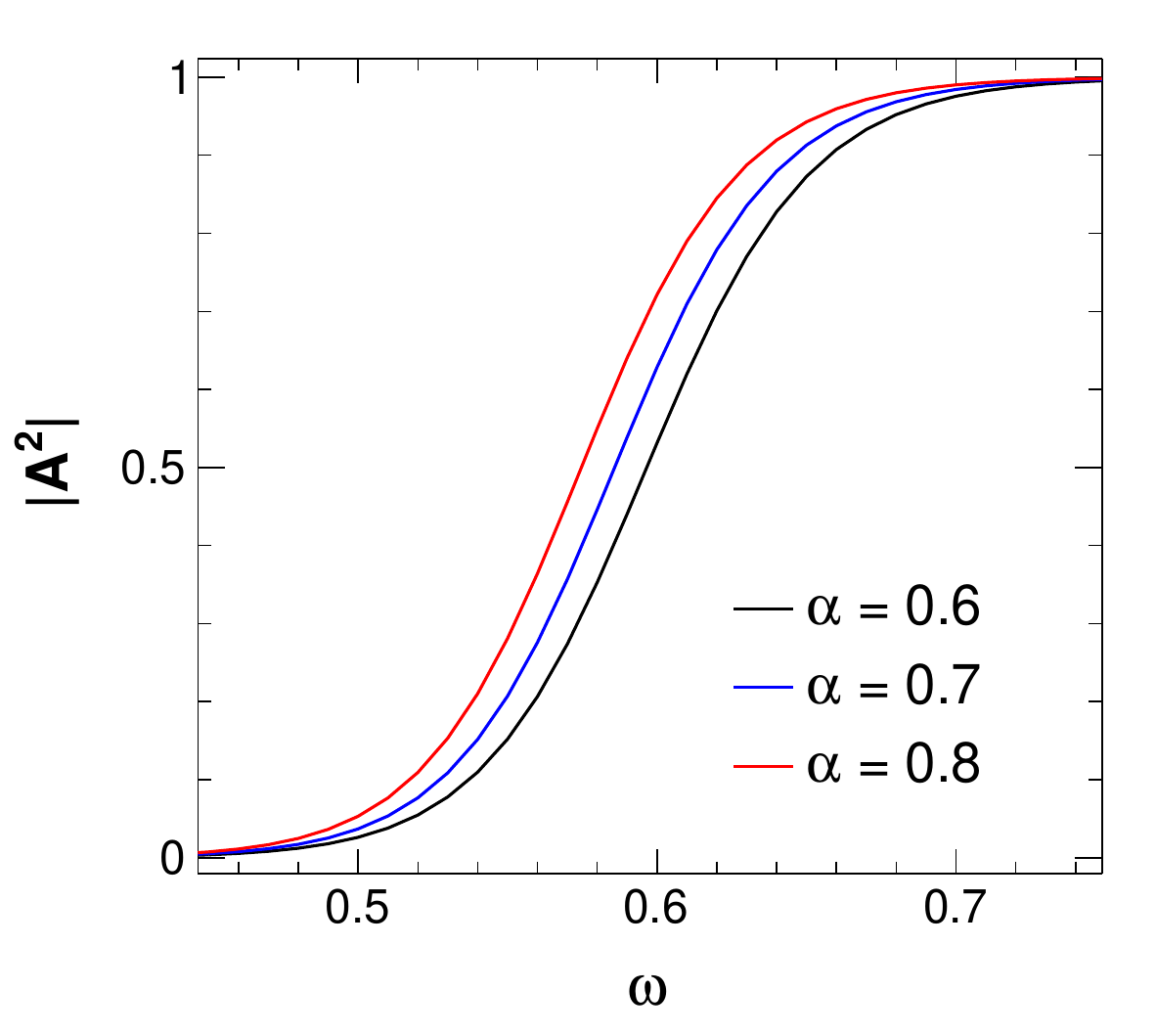}
\vspace{-0.2cm}
\caption{Greybody factor versus quasinormal mode frequency $\omega$ for 
different values of the model parameter $\alpha$. The left plot is for 
multipole $l=1$ and the right plot is for multipole $l=2$. Here $\beta=0.1$, 
$q=1$ and $a=0.01$ used.}
\label{091}
\end{figure}

In this section, we analyse the transmission coefficient or the greybody 
factor for the black hole \cite{27,49,50,51,53,54}. The greybody factor tells 
us about the amount of radiation near the black hole that is trapped or 
reflected by the black hole. The greybody factor gives an idea of the 
probability for the outwards traveling wave to reach an observer at infinity, 
without getting absorbed or the probability of an incoming wave getting 
absorbed by the black hole. The study of reflection coefficient and 
transmission coefficient (greybody factor) is not new, a lot of 
researchers have already studied these coefficients in various scenarios. The 
greybody factor for the scalar perturbation in the Bardeen-de Sitter black 
hole setup has been 
studied in detail in \cite{48-3}. More recently, the greybody factor for the 
Bardeen-de Sitter black hole has been studied for electromagnetic as well 
as gravitational perturbations \cite{49}. The reflection and transmission of 
the wave hitting the barrier potential can be of the following form \cite{49}:
\begin{align}
\psi(x) & = T(\omega) \exp (-i \omega x), \;\; x \rightarrow -\infty,
\label{d1}\\[8pt]
\psi(x) & =  \exp (-i \omega x) + R(\omega) \exp(i \omega x),\;\; x \rightarrow +\infty,
\label{d2}
\end{align}
where $R(\omega)$ and $T(\omega)$ are reflection and transmission coefficients
respectively, and functions of $\omega$. The condition of the conservation of
probability demands that $|R|^2 + |T|^2=1.$ These two coefficients can be 
obtained from the WKB approximation approach as \cite{27,49,50,51,53,54}.
\begin{align}
|R(\omega)|^2 & =\frac{1}{1+\exp(-2\pi i \delta)},
\label{d4}\\[8pt]
|T(\omega)|^2 & = \frac{1}{1+\exp(2\pi i \delta)},
\label{d5}
\end{align}
where $\delta$ parameter can be determined as \cite{27,49},
\begin{equation}
\delta=\frac{i (\omega^2 -V_0)}{\sqrt{-2V_0^{''}}}-\Lambda_j.
\label{d6}
\end{equation}
Expression for the WKB correction terms $\Lambda_j$ can be found in 
Ref.\ \cite{48}. Here $V_0$ represents the effective potential maximum and the 
double dash is for the double derivative with respect to $x$. We plot the 
greybody factors (square of the transmission coefficient) versus $\omega$ for 
different values of the model parameters in Figures \ref{07}, \ref{08}, 
\ref{09} and \ref{091}. Figure \ref{07} shows the greybody factor versus 
$\omega$ for varying values of the string parameter $a$ with $q=1$ and 
$\beta=0.01$ taking the multipole $l=1$ (left) and $l=2$ (right). It is clear 
that for higher values of $a$, $|T(\omega)|^2$ value slightly increases 
starting from the smaller frequency $\omega$. The changes become more visible 
for $l=2$ as shown in the figure (right plot). An increase in the value of the 
transmission coefficient suggests that there is less scattering of the wave 
back from the barrier with an increasing value of $a$. However, at very small 
and large values of $w$, the parameter $a$ does not have any noticeable effect. 
Moreover, we see that although the pattern of variation of greybody factor of 
the black hole is similar to the Schwarzschild black hole, the pattern is 
shifted to low frequency side for $l=l$, whereas it is shifted to the high 
frequency side for
higher $l$ values in comparison to that for the Schwarzschild black hole.   
Figure \ref{08} shows the same plots but for the varying charge $q$ with 
$a=0.01$ keeping the same $\beta$ value. Here, we see that with an increase in 
charge, the greybody factor decreases and thus scattering increases. The effect
gets amplified on increasing multipole $l$ from 1 to 2. Figure \ref{09} shows a 
similar plot for varying $\beta$ with $q=1$ keeping the same value of $a$. Here 
it is seen that higher $\beta$ values result in higher values of 
greybody factor and hence less scattering. Figure \ref{091} shows the greybody 
plots for variation in $\alpha$ values. Similar to the previous cases, here 
also greybody factor increases with increasing $\alpha$ values.\\

The absorption cross-section of the black hole corresponding to 
the transmission coefficient can be computed using the relation \cite{49},
\begin{equation}
\sigma=\Sigma_{l}\,\frac{\pi (2l+1)}{\omega^2}\,|T_l(\omega)|^2.
\label{d7}
\end{equation}
\begin{figure}[h!]
\centerline{
\includegraphics[scale=0.34]{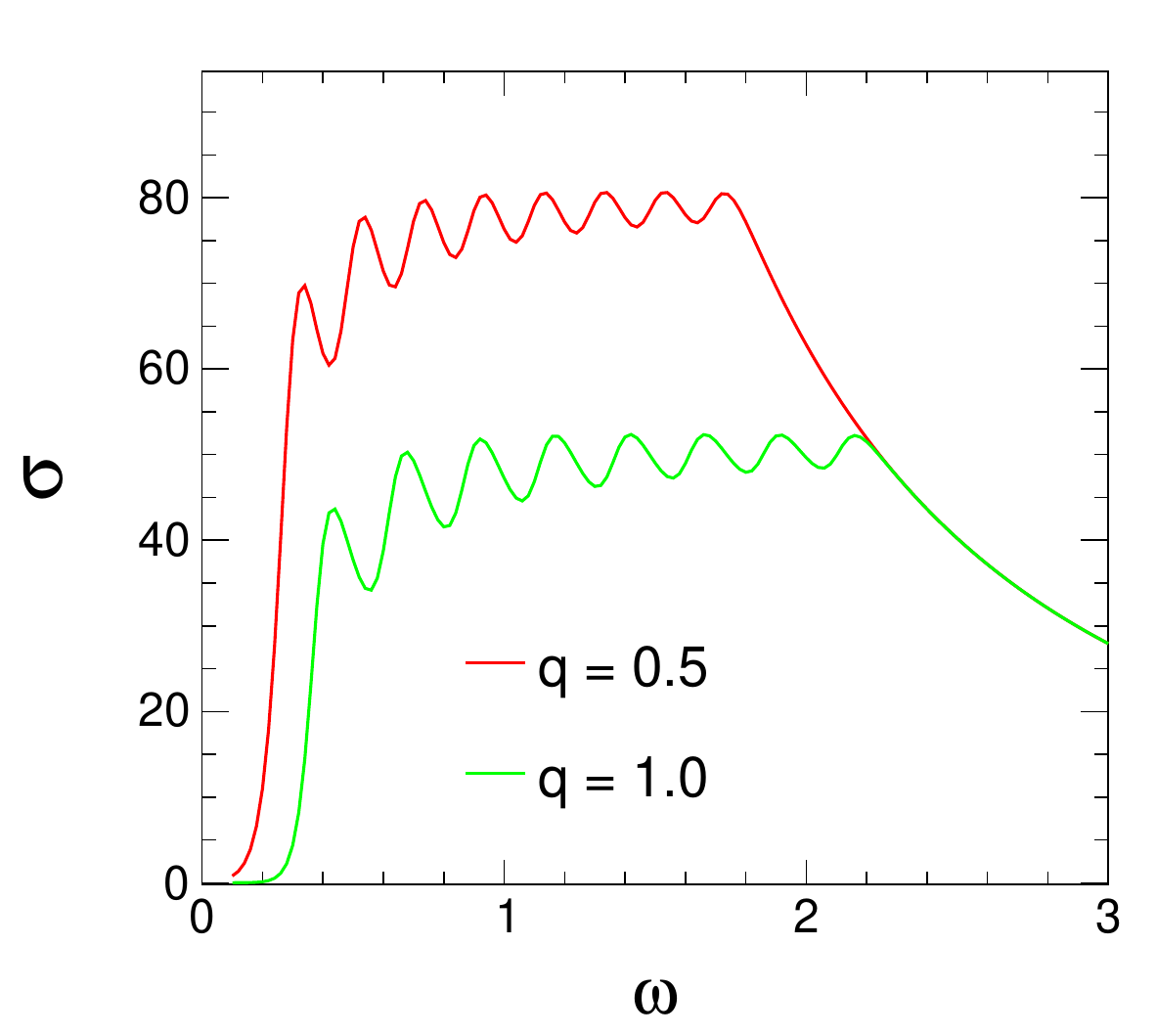}\hspace{0.5cm}
\includegraphics[scale=0.34]{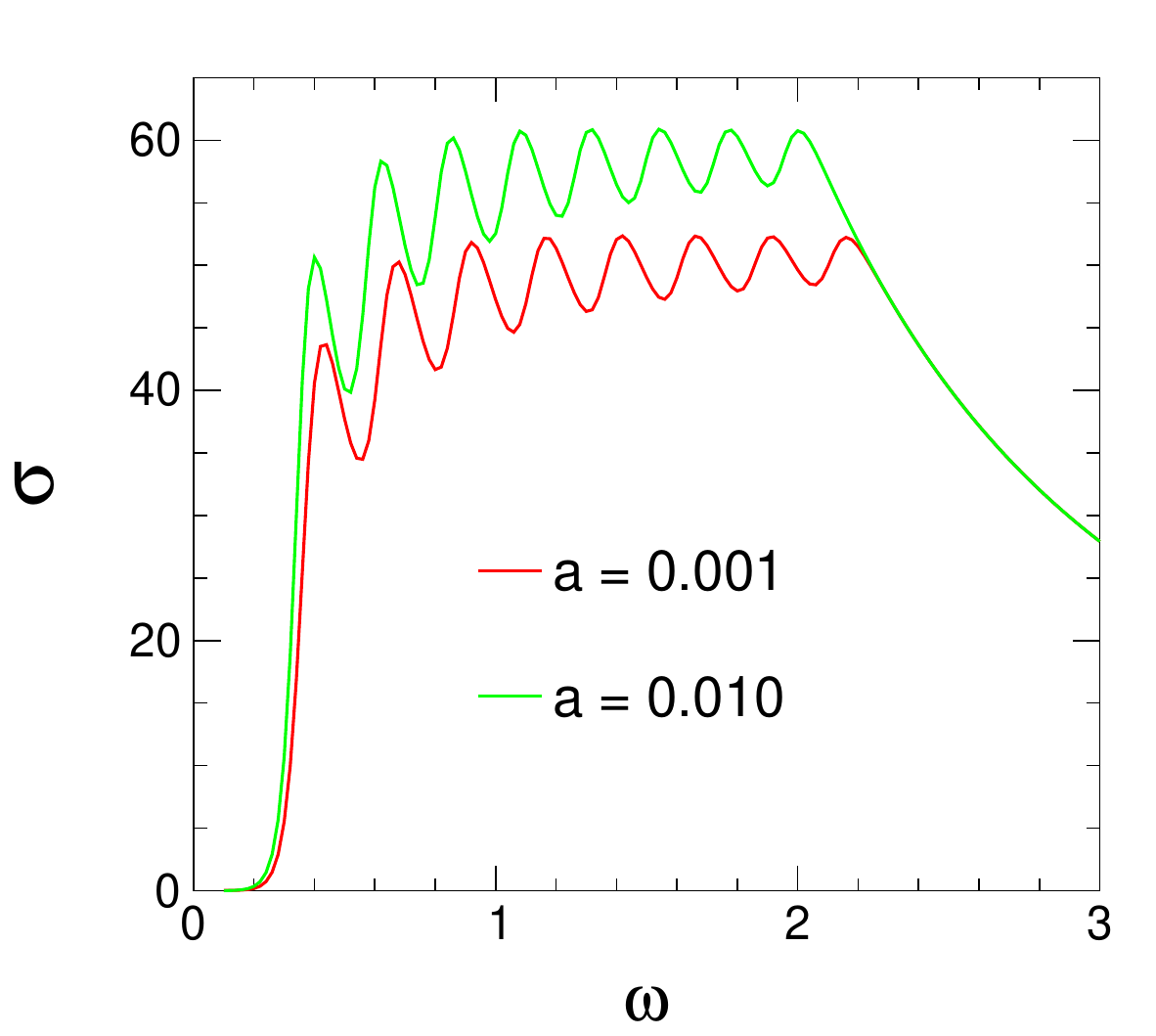}}\vspace{0.3cm}
\centerline{
\includegraphics[scale=0.34]{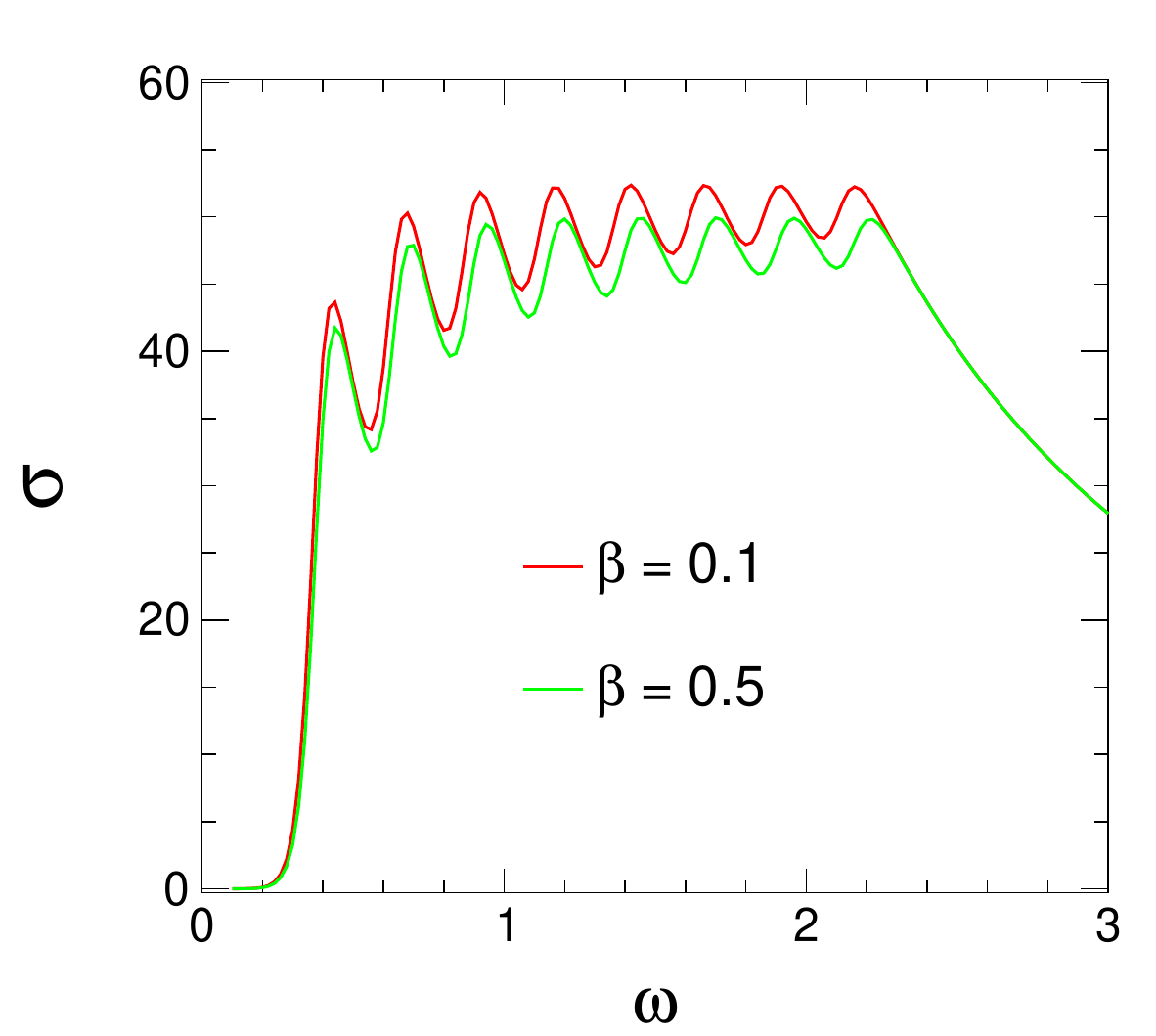}\hspace{0.5cm}
\includegraphics[scale=0.34]{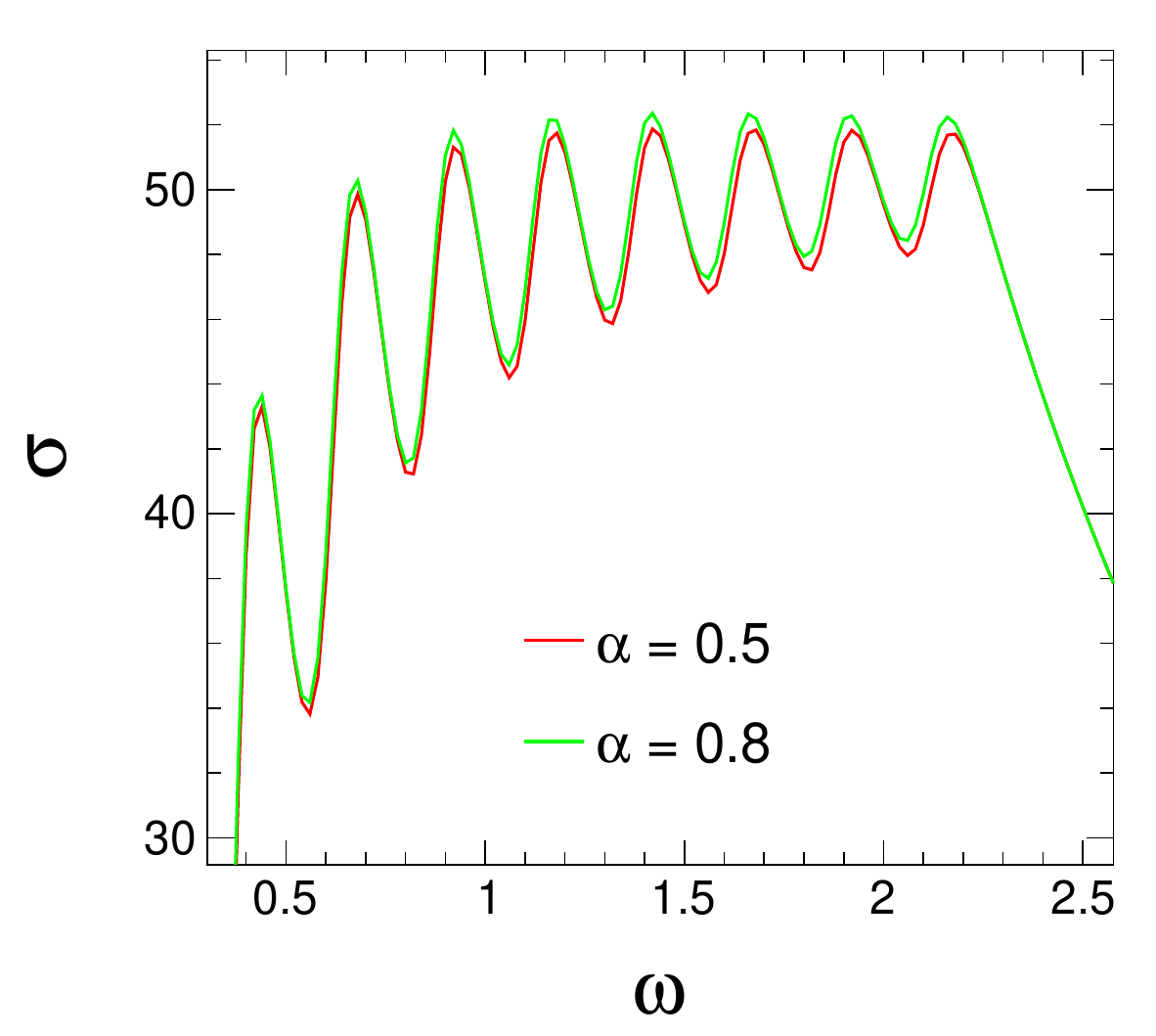}}
\vspace{-0.2cm}
\caption{Absorption cross-section $\sigma$ versus $\omega$ for two different
values of $q$ (first plot), $a$ (second plot), $\beta$ (third plot) and
$\alpha$ (fourth plot) with the required set of remaining three parameters' 
values taken from the set: $q=1$, $\beta=0.1$, $a=0.001$ and $\beta = 0.80$.}
\label{10}
\end{figure}
In Figure \ref{10}, we show the total absorption cross-section for the black 
hole as a function of $\omega$ with values of the model parameters as shown 
in the caption. Here we consider the summation of multipoles $l$ upto 8 
terms in the expression. It is clear that the cross-section value increases 
abruptly with $\omega$ to a maximum value and then saturates for some range of
$\omega$ and finally decreases gradually for higher $\omega$ values. The plot 
can be superficially divided into three regions. The first region is the 
increasing region, which corresponds with the increasing transmission 
coefficient plots that we have shown. The second part is the oscillating part 
due to many multimodes considered in finding $\sigma$. The third region shows 
a power-law type falloff which is due to saturation of transmission coefficient 
value to 1, and hence when $\omega$ is higher valued, then 
$\sigma \propto \frac{1}{\omega^2}$. Moreover, it is seen that $\sigma$ 
increases with increasing values of $a$ and $\alpha$, while it decreases with 
increasing $q$ and $\beta$ values. The effects of $q$ and $a$ on 
$\sigma$ are appreciable with the $q$ domination as the most significant.

\section{Conclusion} \label{sec6}
In this work, we modify the Rastall gravity theory and compute the 
corresponding black hole metric with a cloud of strings field surrounding 
the black hole. The effects of the parameters of the theory on the horizon 
of the black hole have been analysed. The string parameter and the gravity 
model parameters influence the outermost horizon while the charge $q$ impacts 
the inner horizons. We then compute the scalar QNMs for the black hole using 
the 6th order WKB approximation method and observe the dependence of the 
amplitude and damping on various parameters. It is observed that the amplitude 
decreases while the damping increases (for $l=1$ there is a slight change in
the pattern) with an increase in the string parameter $a$. It is seen that with 
an increase in $q$, the amplitude increases, whereas the damping increases at 
the beginning but later decreases. For $\beta$, it is observed that the 
amplitude decreases very minutely, while damping decreases noticeably with 
increasing $\beta$ values. Moreover, it found that for the physical 
consistency the values of the parameter $\alpha$ should be greater than $0.6$. 
Thus for higher values of the $\alpha$ parameter, the amplitude and damping of 
QNMs are seen to be decreased with increasing $\alpha$ values 
(see Table \ref{tab01}). We estimate the approximate errors associated with 
the WKB calculations as shown in the Table \ref{tab01}. 

Further, the convergence of QNMs for various orders of the WKB method has been 
checked. We plot the quality factor of the emission of QNMs to estimate the 
strength of amplitude over damping. Then we plot the temperature of the black 
hole versus the horizon radius for different model parameters. In all the 
cases, the temperature plots show a peak for smaller values of horizon radius 
and then go towards negative values for very small and higher values of the 
horizon radius, indicating ultracold black hole formation possibility in such 
cases. We compute the greybody factors associated with the black hole metric 
and plot the same versus frequency $\omega$ for different model parameters. We 
see that for larger values of $a$, the transmittance increases more rapidly for 
$l=1$. For smaller values of $q$, the transmittance is more. It is also seen 
that transmittance increases with increase in $\beta$ and $\alpha$ values. In 
all four cases, it is seen that the saturation of transmittance value 
occurs more quickly for $l=1$ compared to $l=2$ multipole value. We also 
compute the total absorption cross-section ($\sigma$) and plotted it with 
respect to $\omega$ for variations in different model parameters. It is found 
that the higher the parameters $a$ and $\alpha$ are, the more the 
absorption cross-section. However, the reverse is observed for the cases of 
$q$ and $\beta$ parameters. The parameters $q$ and $a$ affect significantly the 
absorption cross-section.

It needs to be mentioned that efforts are going on to improve the efficiency 
and the sensitivity of the existing GW detectors. 
Moreover, new detectors of GWs like LISA \cite{lisa1,lisa2} and the Einstein 
Telescope \cite{et} will come up with a significant improvement in detection 
sensitivity, which will allow one to constrain our model and other theories of 
gravity very effectively and help to eliminate redundant propositions. We 
believe that more work needs to be done in this direction and with time, as we 
have more and more accurate information regarding different properties of 
black holes, especially of the QNMs, it will be more convenient to validate 
various MTGs, which is one of the sought after intentions in this arena. As a 
future scope, we can study the electromagnetic and gravitational QNMs of the 
black hole for our particular setup. Further, the black hole shadow data can 
also be used to perform a constraining study on various parameters of the 
theory. Furthermore, one can also perform the classical tests of GR, namely 
precession of planetary orbits, Shapiro time delay and gravitational bending 
of light to suggest some constraints on various parameters of the theory. 
Other possible scopes of the obtained black hole solution include 
gravitational lensing study and accretion physics. These are some of the 
possible future prospects of the work.

\section*{Acknowledgements} UDG is thankful to the Inter-University Centre for
Astronomy and Astrophysics (IUCAA), Pune, India for awarding the Visiting 
Associateship of the institute.

\end{document}